\documentclass[useAMS,usenatbib]{mn2e}

\usepackage {graphicx}
\def\aj{\,{AJ}}
\def\apj{\,{\rm ApJ}}
\def\apjl{\,{\rm ApJL}}
\def\apjs{\,{\rm ApJS}}
\def\pasp{\,{\rm PASP}}

\def\araa{\,{\rm ARAA}}
\def\aap{\,{\rm A\&A}}
\def\nat{\,{\rm nature}}
\def\mnras{\,{\rm MNRAS}}

\title[The Morphology-Density Relation in the Sloan Digital Sky Survey]{The Morphology-Density Relation in the SDSS}
\author[Tomotsugu Goto  et al.]{Tomotsugu Goto$^{1,2}$\thanks{E-mail:
yohnis@icrr.u-tokyo.ac.jp}, Chisato
Yamauchi$^{3}$, Yutaka Fujita$^{3}$, Sadanori Okamura$^{2}$,
\newauthor Maki Sekiguchi$^{1}$, Ian Smail$^{4}$, Mariangela Bernardi,$^{5}$
and Percy L. Gomez$^{5}$
\\
$^{1}$Institute for Cosmic Ray Research, University of
Tokyo, Kashiwanoha, Kashiwa, Chiba 277-0882, Japan\\
$^{2}$Department of Astronomy, Graduate School of Science, The University of Tokyo,
Hongo 7-3-1, Bunkyo-ku, Tokyo 113-0033, Japan\\
$^{3}$National Astronomical Observatory, 2-21-1 Osawa, Mitaka, Tokyo
181-8588,Japan\\
$^{4}$Department of Physics, University of Durham, South Road, Durham
DH1 3LE, UK\\
$^{5}$Department of Physics, Carnegie Mellon University, 
5000 Forbes Avenue, Pittsburgh, PA 15213, USA\\
}
\begin{document}


\pagerange{\pageref{firstpage}--\pageref{lastpage}} \pubyear{2002}

\maketitle

\label{firstpage}

\begin{abstract}
 
%
%

 We have studied the morphology-density relation and
 morphology--cluster-centric-radius relation using a volume limited
 sample (0.05$<z<$0.1, $Mr^*<-$20.5) of the Sloan Digital Sky Survey
 (SDSS) data. Major 
 improvements compared with previous work are; (i) automated galaxy
 morphology classification capable to separate galaxies into four types,
 (ii) three 
 dimensional local galaxy density 
 estimation, (iii) the extension of the morphology-density relation into
 the field region. 
 We found that the morphology-density and
 morphology--cluster-centric-radius relation in the 
 SDSS data for both of our automated morphological classifiers, $Cin$ and
 $Tauto$, as fractions of early-type galaxies increase and late-type
 galaxies decrease toward increasing local galaxy density. 
    In addition, we found there are two characteristic changes in both the
 morphology-density and the morphology-radius relations, suggesting two
 different mechanisms are responsible for the relations.
  In the sparsest regions (below 1 Mpc$^{-2}$ or outside of 1 virial
 radius), both relations  become less noticeable, suggesting the
 responsible physical mechanisms for galaxy morphological change require
 denser environment.  
   In the intermediate density regions (density between 1 and 6
 Mpc$^{-2}$ or virial radius between 0.3 and 1), intermediate-type
 fractions increase 
 toward denser regions, whereas late-disc fractions
 decrease. Considering the median size of intermediate-type galaxies are
 smaller than 
 that of late-disc galaxies, we propose that 
  the mechanism is likely to stop star formation in
 late-disc galaxies, eventually turning them into intermediate-type
  galaxies after their  outer discs and spiral arms become 
 invisible as stars die. For example, ram-pressure stripping is
 one of the candidate mechanisms. 
   In the densest regions (above 6 Mpc$^{-2}$ or inside of
 0.3 virial radius), intermediate-type fractions decrease radically and early-type
 fractions increase in turn. This is a contrasting result to that in
 intermediate regions and it suggests that yet another mechanism is more
 responsible for the morphological change in these regions.

%

 We also compared the morphology-density relation from the SDSS
 (0.01$<z<$0.054) with that of the MORPHS data ($z\sim$0.5). Two
 relations lie on top of each other, suggesting that the
 morphology-density relation was already established at $z\sim$0.5 as is
 in the present universe. A slight sign of an excess elliptical/S0 fraction in
 the SDSS data in dense regions might be suggesting the additional formation
 of elliptical/S0 galaxies in the cluster core regions between $z=$0.5 and $z=$0.05.

\end{abstract}


\section{Introduction}\label{intro}

 Morphological types of galaxies are one of the most basic properties
 and thus have been studied since the beginning of the extragalactic
 astronomy (e.g., Roberts \& Haynes 1994). However, it is still not very
 well understood where this 
 diversity stems from. The existence of a correlation between galaxy
 morphology and local environment is a remarkable feature of galaxy population.
 Dressler (1980) studied 55 nearby galaxy clusters and found that
 fractions of elliptical 
 galaxies increase and that of spiral galaxies decrease with
 increasing local galaxy density in all clusters.
  The discovery left a great impact on
 astronomical community since it indicates that physical mechanisms that
 depend on environment of each galaxy mainly affect the final
 configuration of stellar component. Further observational constraints
 on the formation and evolution of galaxies were obtained by extending
 the analysis of the morphology-density relation to group of galaxies in
 the general field. Postman \& Geller (1984) extended morphology study
 to groups using the data from the CfA Redshift Survey (Huchra et
 al. 1983). The relation was completely consistent with Dressler
 (1980). At low densities, population fractions seemed to be independent
 of density below galaxy density $\sim$5 Mpc$^{-3}$. At high density,
 the elliptical fraction increased steeply above $\sim$3000 galaxies
 Mpc$^{-3}$. 
  Therefore, they proposed the morphology-density relation have three
 distinct regions or two breaks which are related to the timescale of
 different physical mechanisms acting upon the galaxy populations.
 The morphology-density relation is also observed in X-ray selected poor groups.
  Tran et al. (2001) studied six nearby X-ray detected, poor groups and
 found that the fraction of bulge-dominated galaxies in groups decreases
 with increasing radius, similar to the morphology-density relations in clusters.
 Helsdon \& Ponman
 (2002) observed the morphology-density relation in groups and
 found that X-ray bright groups have a lower spiral fraction. 
   The opposite results on groups came from Whitmore et al. (1995). They analyzed
 the morphology-density relation in groups of galaxies by carefully
 removing cluster galaxies from their analysis and found that the
 relation is very weak or non-existent in groups. 

  It is important to note that Whitmore et al. (1991,1993) re-analyzed
 the 55 nearby clusters 
 (Dressler 1980) and argued that the morphology-density relation
 reflects a more fundamental morphology-radius relation; the
 correlation between morphology and cluster centric radius seems tighter
 than the morphology-density relation. Although this assertion is still
 controversial, it may have a significant implication for underlying
 physical mechanisms. 
 
 Later the relation between morphology and density was traced back to
 higher redshift. 
 Dressler et al. (1997) studied 10 high redshift clusters at $z\sim$0.5
 and found that the morphology-density relation is strong for centrally
 concentrated clusters. However, the relation was nearly absent for less
 concentrated or irregular clusters. They also found that S0 fractions
 are much smaller than in nearby clusters, suggesting that S0 galaxies
 are created fairly recently ($z\leq$0.5). 
 At $z=0.4$, Treu et al. (2003) found that the fraction of early-type
 galaxies declines steeply from the cluster center to 1 Mpc radius using
 the cluster Cl0024+16. 
 Fasano et al. (2000) studied nine clusters at intermediate redshift 
 (0.1$\leq z\leq$0.25) and compared them with local (Dressler 1980) and
 high redshift clusters (Dressler et al. 1997). They found that
 morphology-density relation in  high elliptical concentration clusters,
 but not in low elliptical concentration clusters.
 The finding is consistent with the results of Dressler et
 al. (1997). Considering that 
 low redshift clusters have morphology-density relation regardless of
 the concentration of clusters, they suggested that spiral to S0 transition
 happened fairly recently (last 1-2 Gyr). They also plotted morphological
 fraction as a function of redshift and found that S0 fraction decreases
 with increasing redshift, whereas spiral fraction increases with
 redshift.  

 Hashimoto et al. (1999) used data from Las Campanas Redshift Survey
 (LCRS; Shectman et al. 1996) to study the concentration-density
 relation. They found that the ratio of high and low concentrated
 galaxies decreases smoothly with decreasing density.    
 Dominguez et al. (2001) analyzed nearby clusters with X-ray and found
 that mechanisms of global nature (X-ray mass density) dominate in high
 density environments, 
 namely the virialized regions of clusters, while local galaxy density
 is the relevant parameter in the outskirts where the influence of
 cluster as a whole is relatively small compared to local effects.  
 Dominguez et al. (2002) studied groups in 2dF Galaxy Group Catalog
 using PCA analysis of spectra as a galaxy classification and local galaxy
 density from redshift space as a measure of galaxy environment. 
 They found that both morphology-density relation and
 morphology-group-centric radius relation is clearly seen in high mass
 ($Mv\geq$10$^{13.5}M_{\odot}$) groups, but neither relation holds true for
 low mass ($Mv<$10$^{13.5}M_{\odot}$) groups. 
 These three studies made innovative step in terms of an analysis
 method, using automated morphological classification  and three dimensional density estimation.

   Various physical mechanisms have been proposed to explain the
  morphology-density relation.
  Possible causes include ram-pressure stripping of gas (Gunn \& Gott 1972; Farouki
 \& Shapiro 1980; Kent 1981; Fujita \& Nagashima 1999;
  Abadi, Moore \& Bower 1999; Quilis, Moore \& Bower 
 2000), 
  galaxy harassment via high speed impulsive
  encounters (Moore et al. 1996, 1999; Fujita 1998), cluster 
 tidal forces (Byrd \& Valtonen 1990; Valluri 1993; Fujita 1998) which distort
 galaxies as they come close to the centre, interaction/merging of
 galaxies (Icke 1985; Lavery \& Henry 1988, Mamon 1992; Makino \& Hut
  1997; Bekki 1998;  Finoguenov et al. 2003a), and removal \& consumption of the gas due to the cluster environment (Larson, Tinsley \& Caldwell 1980; Bekki et al. 2002;  Finoguenov et al. 2003b).  
 Shioya et al. (2002) showed that the
  truncation of star formation can  explain the decrease of S0 with increasing redshift. 
 Although these processes are all plausible, the effects provided by initial
  condition on galaxy formation could be also important. Since field
  galaxies have different population ratio compared with clusters (Goto
  et al. 2002a), a change in
  infalling rate of field galaxies into clusters affects population
  ratio of various galaxy types (Kodama et al. 2001).
 Unfortunately, there exists little evidence demonstrating that any one
  of these processes is actually responsible for driving galaxy
  evolution. Most of these processes act over an extended period of
  time, while observations at a certain redshift cannot easily provide
  the detailed information that is needed to elucidate subtle and
  complicated processes.  

 To extract useful information from observational data, it is necessary
 to have detailed theoretical predictions.
 In recent years, due to the progress of computer technologies,
 it is becoming possible to simulate the morphology-density relations by
 combining semi-analytic modeling with N-body simulations of cluster
 formation.  
 Okamoto \& Nagashima (2001) simulated the morphology-density relation
 using a merger-driven bulge formation model. They found that early-type
 fractions are well re-produced, but there remained a discrepancy on intermediate-type
 fractions.  Diaferio et al. (2001) also assumed that the
 morphologies of cluster galaxies are determined solely by their merging
 histories in the simulation. They used bulge-to-disc ratio to classify
 galaxy types and compared the cluster-centric radial distribution with
 those derived from CNOC1 sample (Yee,Ellingson, \& Carlberg 1996). They
 found excellent agreement for bulge 
 dominated galaxies, but simulated clusters contained too few galaxies
 of intermediate bulge-to-disc ratio.  Springel et al. (2001) used a
 phenomenological simulation to predict the morphology-radius relation
 and compared it with Whitmore et al.(1993). Their morphological
 modeling is based on the merging history of galaxies. They found an
 excellent agreement with early-type galaxy fractions, and some
 deficiency of intermediate-type galaxies in the core of the cluster. Benson et
 al. (2001) combined  their N-body simulation with
 a semi-analytic model (Cole et al. 2000) to trace the
 time evolution of the morphology-density relation. Interestingly, they
 found that a strong morphology-density relation was well established at
 $z=$1. The relation was qualitatively similar to that at $z=$0. E/S0
 galaxies are treated as one population in their simulation.
 Three of above simulations
 suggest that (i) early-type fractions are consistent with the merging
 origin; (ii) however the deficit of intermediate-type galaxies shows
 that processes other 
 than major-merger might be important 
 for intermediate-type creation. Therefore, more than one mechanism might be required to
 fully explain the morphology-density relation. These suggestions might be
 consistent with observational results from  Dominguez et al. (2001),
 who found two different key  parameters in cluster centre and outskirts
 separately.  
  
  In the previous analysis of the morphology-density relation from
  observations, there have been two major difficulties; eye-based
  morphological classification and the density estimate from two-dimensional imaging data. 
   Although it is an excellent tool to
 classify galaxies, manual selection could potentially have unknown
 biases (see Lahav et al. 1995; Fabricant et al. 2000).  A machine based, automated classification 
 would better control biases and would allow a reliable determination of
 the completeness and false positive rate.
  Measuring local galaxy density from imaging data requires statistical
  background subtraction, which automatically invites relatively large uncertainty
  associated with itself. 
  Furthermore the deeper imaging data
  , the bigger the correction. Therefore, three dimensional density
  determination from redshift data is preferred. 
    With the advent of the Sloan Digital Sky Survey (SDSS; York et al. 2000),
  which is an imaging and spectroscopic survey of 10,000 deg$^2$ of the
  sky, we now have the opportunity to overcome these
  limitations. Several of the largest cluster catalog are compiled using the
  SDSS data (Annis et al. 1999; Kim et al. 2002; Goto et al. 2002b).
  The
  CCD imaging of the SDSS allows us to estimate morphologies of galaxies
  in an automated way (Yamauchi et al. 2003). Three dimensional density
  can be estimated from the redshift information (e.g., Hogg et
  al. 2003).  Due to the large area 
  coverage of the SDSS, we are able to probe the morphology-density
  relation from cluster core regions to the field region without
  combining multiple data sets with inhomogeneous characteristics.
    The purpose of this paper is as follows. We aim to confirm or
  disprove the morphology-density relation using the automated morphology
  and three dimensional density from the SDSS data. We also re-analyze
  the MORPHS data ($z\sim$0.5)
  using an automated morphology (Smail et al. 1997).
  By comparing them to the SDSS, we
  try to observe the evolution of the morphology-density relation.   
 Final goal of our investigation is to shed some light on the origin
  of the morphology-density relation.

 The paper is organized as follows: In
 Section \ref{data}, we describe the SDSS data. 
 In Section \ref{analysis}, we explain automated morphological classifications and
 density estimation.
 In Section \ref{results}, we present the results from the SDSS and the
 MORPHS data.
 In Section \ref{discussion}, we discuss the possible caveats and underlying physical
 processes which determines galaxy morphology.  In
 Section \ref{conclusion}, we summarize our work and findings.
   The cosmological parameters adopted throughout this paper are $H_0$=75 km
 s$^{-1}$ Mpc$^{-1}$, and ($\Omega_m$,$\Omega_{\Lambda}$,$\Omega_k$)=(0.3,0.7,0.0).

\section{The SDSS Data}\label{data}
 
 The data we use to study the morphology-density relation are 
 from the Sloan Digital Sky Survey Early Date Release (SDSS EDR;
 Stoughton et al. 2002), which covers $\sim$400 deg$^2$ of the sky  
 (see Fukugita et al. 1996; Gunn et al. 1998;  Lupton
  et al. 1999,2001,2002; York et al. 2000; Eisenstein et al. 2001;
  Hogg et al. 2001; Blanton et al. 2003a; Richards et al. 2002;
 Stoughton et al. 2002; Strauss et   al. 2002; Smith et al. 2002;  Pier et
  al. 2003  for more detail  of the SDSS data.)
 The imaging part of the SDSS observes the sky in five optical bands
 ($u,g,r,i,$ and $z$; Fukugita et al. 1996). 
 Since the SDSS photometric system is not yet finalized, we refer to the
 SDSS photometry presented here as $u^*,g^*,r^*,i^*$ and $z^*$.
 The technical aspects of the SDSS camera are described in Gunn et al.\ (1998).  
 The SDSS spectroscopic survey observes the spectra of essentially all
 galaxies brighter than  $r^*$=17.77.
 The target galaxies are selected from imaging part of the survey
 (Strauss et al. 2002). 
 The spectra are observed using a pair of double fiber-fed
 spectrographs obtaining 640 spectra per exposure of 45 minutes.
 The wavelength coverage of the spectrographs is continuous from about
 3800 \AA{} to 9200 \AA{}, and  the wavelength resolution,
 $\lambda/\delta\lambda$, is 1800. The fiber diameter is 0.2 mm ($3''$ on the sky). 
 Adjacent fibers cannot be located closer than $55''$ on the sky.
 The throughput of the spectrograph will be better than 25\% over  4000
 \AA{} to 8000 \AA{} excluding the loss due to the telescope and
 atmosphere. 
 (See Eisenstein et al. 2001; Strauss et al. 2002 and  Blanton et
 al. 2003a for more detail of the SDSS spectroscopic data).  

 We use galaxies in the redshift range 0.05$<z<$0.1 with a redshift
 confidence of $\geq0.7$ (See Stoughton et
 al. 2002 for more details of the SDSS parameters). 
 The galaxies are limited to 
 $Mr^*<-$20.5, which gives us a volume limited sample with 7938
 galaxies. We correct magnitudes for galactic extinction using reddening map of Schlegel, Finkbeiner \& Davis (1998).
 We use $k$-correction given in Blanton et al. (2003b;v1\_11)  to calculate
 absolute magnitudes.

\section{Analysis}\label{analysis}

\subsection{Morphological Classification}\label{morphology}

\begin{figure}
\includegraphics[scale=0.4]{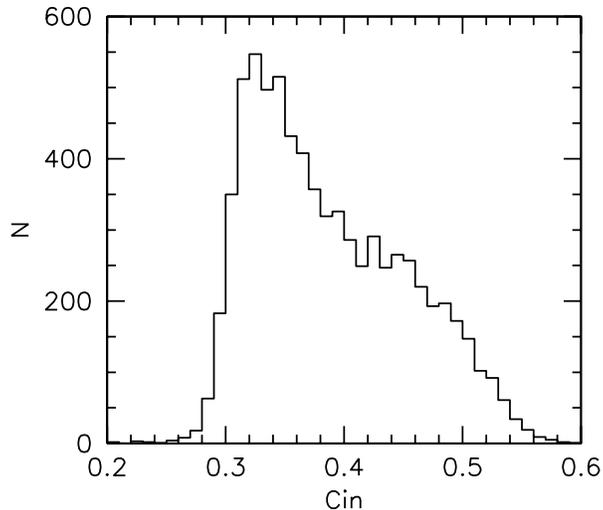}
\caption{
\label{fig:Cin_hist} 
 The distribution of $Cin$ for the volume limited sample of 7938
 galaxies.
}
\end{figure}

\begin{figure}
\includegraphics[scale=0.4]{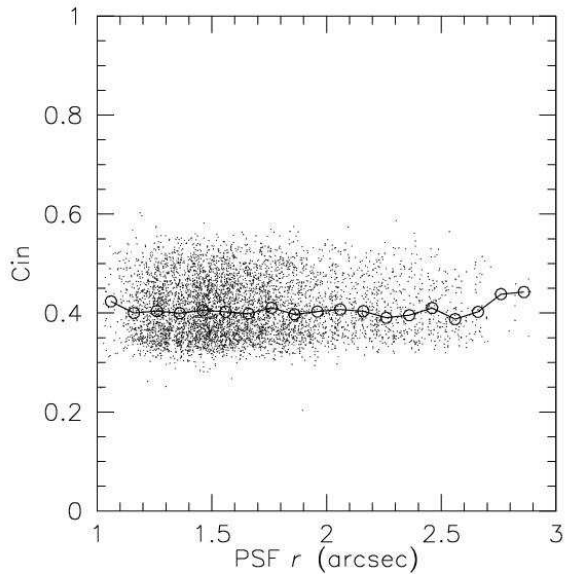}
\caption{
\label{fig:seeing_concent} 
 Seeing dependence of $Cin$. The solid lines show medians. 87\% of our
 sample galaxies have seeing  between 1.2 and  2 arcsec, where seeing
 dependence of $Cin$ is negligible. The seeing is measure in $r$ band. 
}
\end{figure}

\begin{figure}
\includegraphics[scale=0.4]{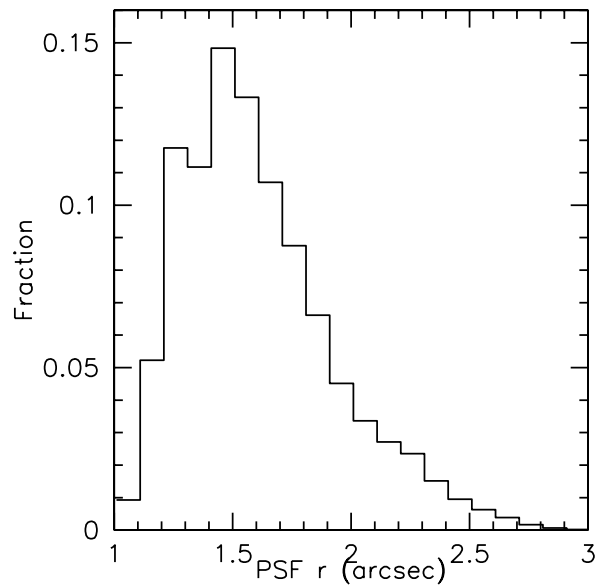}
\caption{
\label{fig:seeing_hist} 
 Distribution of seeing of the SDSS galaxies, measured in $r$ band. 
}
\end{figure}

 We use two different ways of classifying galaxy morphologies. The
 first one is a concentration parameter $Cin$, which is defined as the ratio
 of Petrosian 50\% light radius to Petrosian 90\% light radius. We show
 the distribution of this $Cin$ parameter in Figure \ref{fig:Cin_hist}.
 Shimasaku et al. (2001) and Strateva et al. (2001) showed that this
 $Cin$ parameter correlates well with their eye-classified morphology
 (See Figure 10 of Shimasaku et al. 2001 and Figure 8 of Strateva et
 al. 2001).
 We regard galaxies with $Cin\geq$0.4 as late-type galaxies and ones with
 $Cin<$0.4 as early-type galaxies.  
  The criterion of $Cin$=0.4 is more conservative for late-type
 galaxies. As shown by Shimasaku et al. (2001), $Cin$=0.4 provides
 late-type galaxy sample with little contamination and early-type
 galaxy sample with small contamination.
  The seeing dependence of $Cin$ is presented in figure
 \ref{fig:seeing_concent} for our volume limited sample galaxies. As is
 shown in figure \ref{fig:seeing_hist}, 
 87\% of our sample galaxies have seeing between 1.2 and 2 arcsec where
 the dependence of $Cin$ on seeing size is negligible.

\begin{figure}
\includegraphics[scale=0.4]{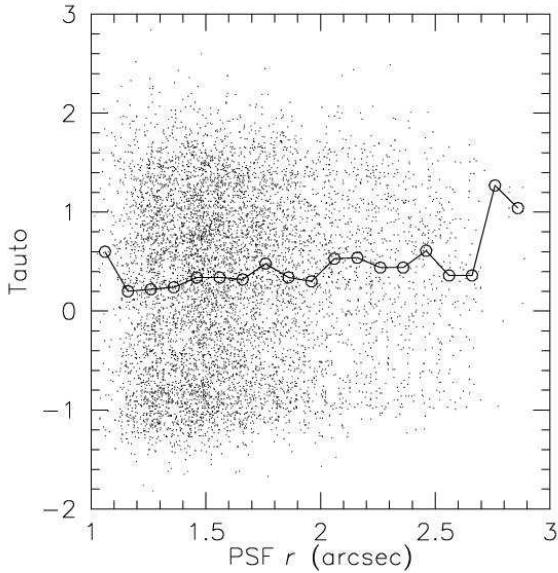}
\caption{
\label{fig:tauto_seeing} 
Seeing dependence of $Tauto$. The solid lines show medians of the distribution. $Tauto$ is essentially independent of seeing size between 1.2 and 2 arcsec, where 87\% of our sample galaxies lie.
}
\end{figure}

\begin{figure}
\includegraphics[scale=0.4]{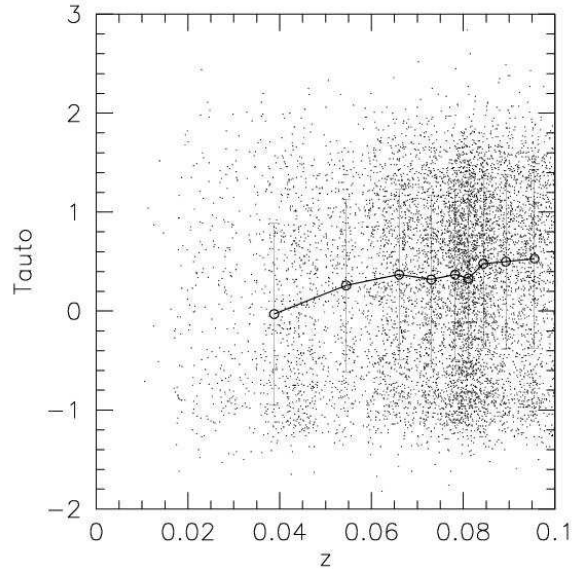}
\caption{
 Redshift dependence of $Tauto$. The solid lines show medians of the
 distribution, which are consistent with constant throughout the redshift
 range we use (0.05$<z<$0.1).
 $Tauto$ shows some deviation at lower redshift ($z<0.04$) since an apparent
 size of a galaxy on the sky radically increases at this low redshift.
}\label{fig:size_z} 
\end{figure}

 The other classification uses morphological parameters measured by
 Yamauchi et al. (2003). We briefly summarize their morphological
 classification. Details on the method and various systematic tests
 including completeness and contamination study are
 given in Yamauchi et al. (2003)\footnote{Note that this work uses a preliminary
 version of $Tauto$ parameter. The final parameter presented in Yamauchi
 et al. (2003) is expected to be improved further.}. The classification method consists of
 two parts. In the first part, concentration index, $Ci$, is calculated as
 the ratio 
 of the Petrosian 50\% light radius and Petrosian 90\% radius
 as is for $Cin$ but the parameter is corrected for elongation of
 galaxies. The elongation correction prevent galaxies with low
 inclination (nearly  edge-on galaxies) from being misclassified as
 early-type galaxies.   
 In the second part, coarseness of galaxies, $Cn$, is calculated as
 the ratio of 
     {\it residuals from the best fit of the galaxy radial profile}
  to {\it difference between the maximum and minimum values of the profile}.
 $Cn$ is sensitive to arm structures of spiral galaxies, and thus larger
 for spiral galaxies with a clear arm structure than  
 galaxies with a smooth radial profile such as ellipticals and S0s. This
 parameter, $Cn$, helps 
 classifying late-type galaxies further into two types of galaxies.
 Finally, $Ci$ and $Cn$ are combined to be a final morphological
 parameter, $Tauto$.  
 Both $Ci$ and $Cn$ are shifted so that their median values become 0.5, and then 
 divided by its standard deviation to be combined to the final parameter
 to classify morphologies as follows.

  \begin{equation} 
 Tauto = Ci(normalized) + Cn(normalized)
  \end{equation}
 $Tauto$ shows better correlation with eye classified morphology than
 $Cin$, as shown in Yamauchi et al. (2003). The correlation
 coefficient with eye-morphology is 0.89.
 Based on the $Tauto$ parameter, we divide galaxies into
 four sub-samples in this study.  We regard galaxies with $Tauto>$1.0 as
 late-disc ($LD$)
 galaxies, 0.1$<Tauto\leq$1.0 as early-disc($ED$),  $-0.8<Tauto\leq$0.1 as
 intermediate-type ($I$, mostly S0s)
 and $Tauto<-0.8$ as early-type ($E$)
 galaxies. Among our sample galaxies, 549 galaxies have eye-classified
 morphologies (Shimasaku et al. 2001; Nakamura et al. 2003).
   In table \ref{tab:completeness}, we quote
 completeness and contamination rate of these four types of galaxies
 classified by $Tauto$, using  eye-classified morphology.  Full
 discussion on contamination and completeness 
 will be given in detail in Yamauchi et al. (2003).    
 As shown in figure \ref{fig:tauto_seeing}, the parameter is
 robust against seeing variance in our volume limited sample galaxies. 
 In figure \ref{fig:size_z}, we plot $Tauto$ against redshift. Medians
 are shown in the solid line. In our redshift range (0.05$<z<$0.1),
 $Tauto$ is essentially independent of redshift related bias.  

\begin{table*}
\begin{center}
\caption{
\label{tab:completeness}
 Completeness and contamination rate of our four sample of galaxies
 classified by $Tauto$ are calculated using eye-classified morphology.
}
\begin{tabular}{llll}
\hline
 Type  & Criteria &Completeness (\%)  & Contamination (\%)\\
\hline
\hline
\noalign{\smallskip}
 Early-type($E$, mostly elliptical) & $Tauto \leq-$0.8         & 70.3 &  28.2   \\
 Intermediate($I$, mostly S0) &        $-$0.8$\leq Tauto<$0.1     & 56.4 &  56.5    \\ 
 Early Disc($ED$, mostly Sa)  &0.1$\leq Tauto<$1.0   & 53.1 &  24.1  \\ 
 Late Disc  ($LD$, mostly Sc)  &  1.0$\leq Tauto$      & 75.0 & 45.9      \\ 
 \hline
\end{tabular}
\end{center}
\end{table*}

\begin{figure}
\centerline{\includegraphics[scale=0.2]{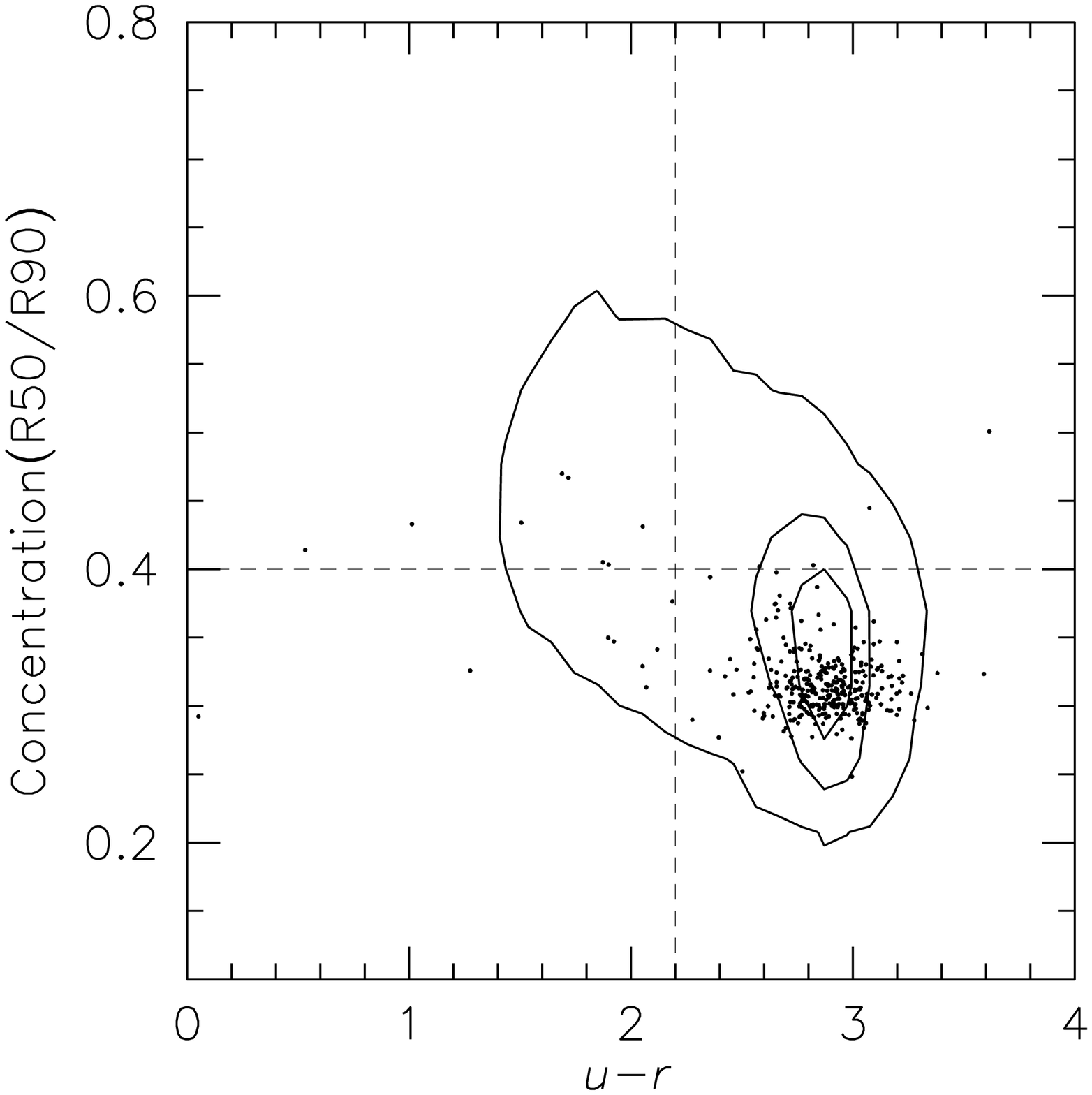}
\includegraphics[scale=0.2]{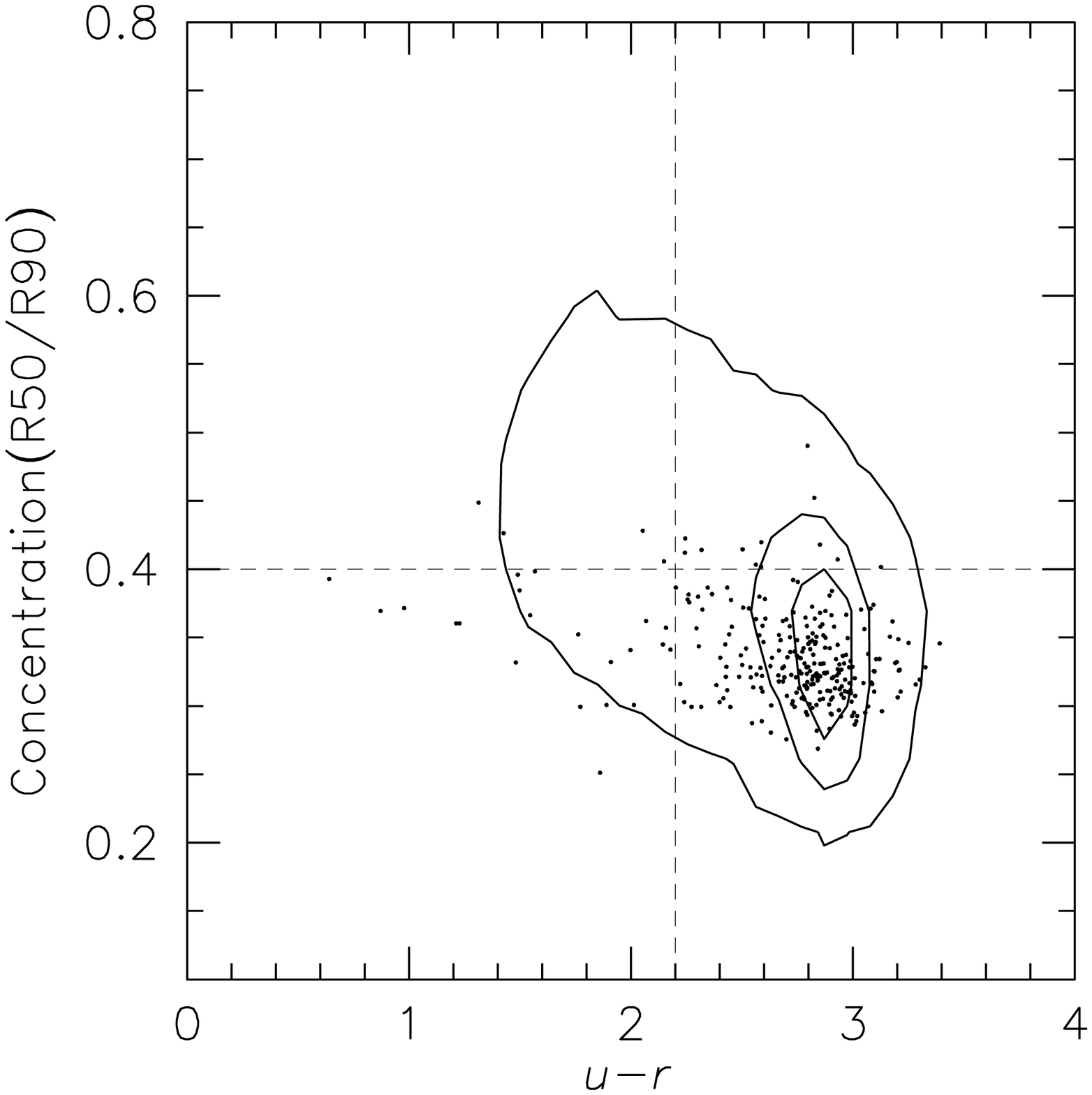}}
\centerline{\includegraphics[scale=0.2]{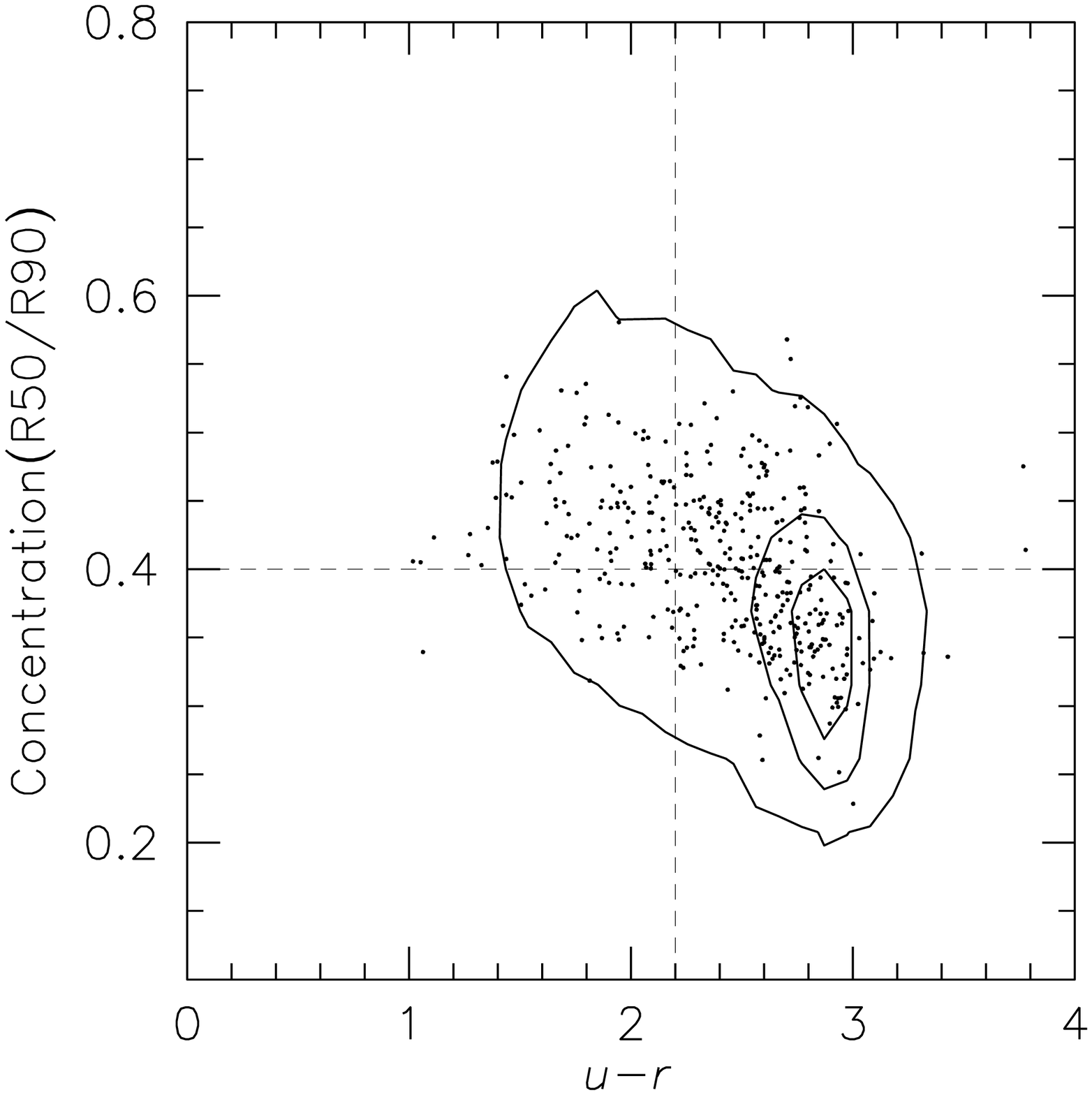}
\includegraphics[scale=0.2]{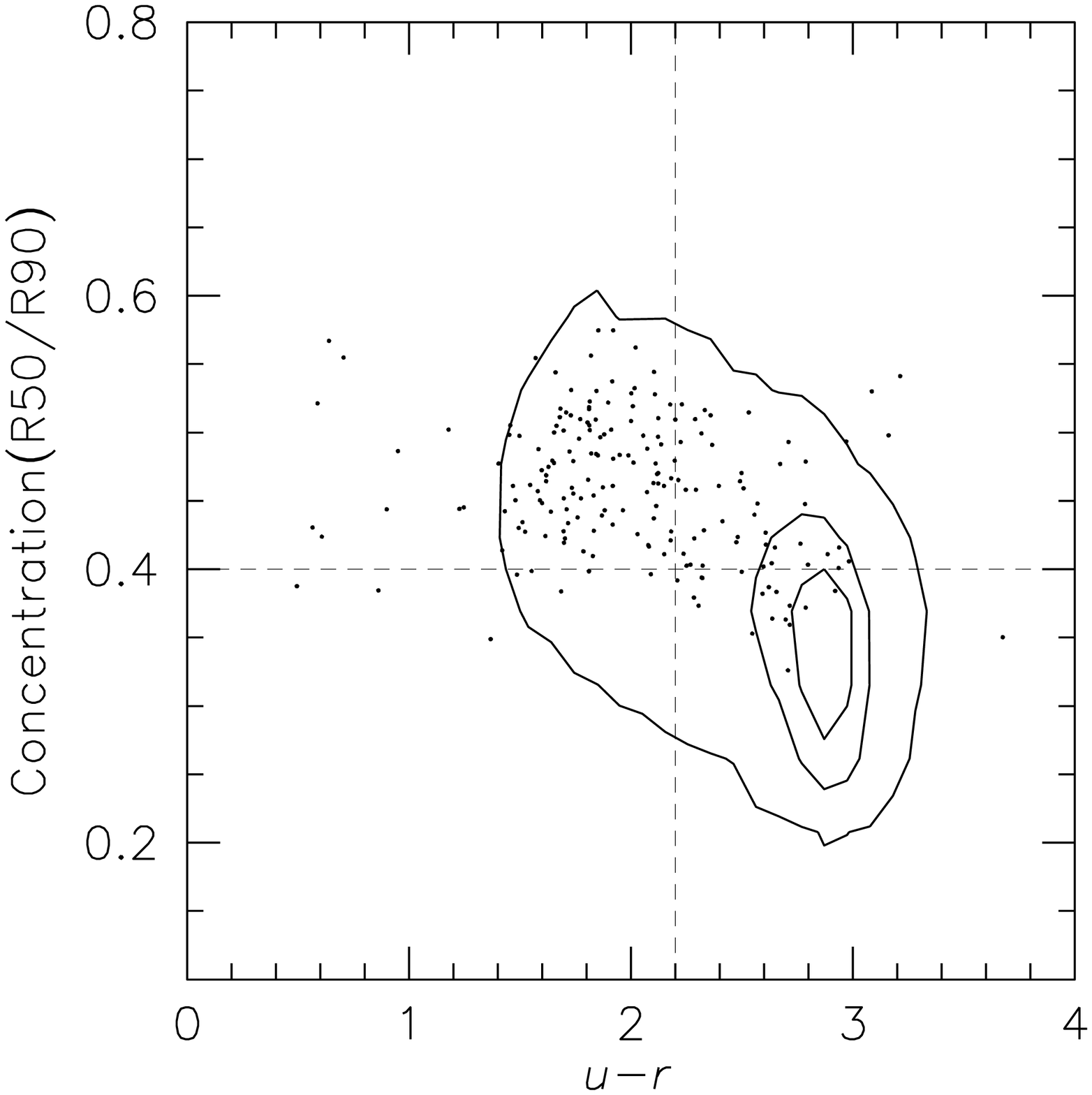}
}
\caption{\label{fig:iskra}
 $Cin$ is plotted against $u-r$. The contours show distribution of all
 galaxies in the volume limited sample.  Points in each
 panel show the distribution of each morphological type of galaxies
 classified by eye (Shimasaku et al. 2001; Nakamura et
 al. 2003). Ellipticals are in the upper left panel. S0, Sa and Sc are
 in the upper right, lower left and lower right panel, respectively.   
}
\end{figure}

\begin{figure}
\centerline{\includegraphics[scale=0.2]{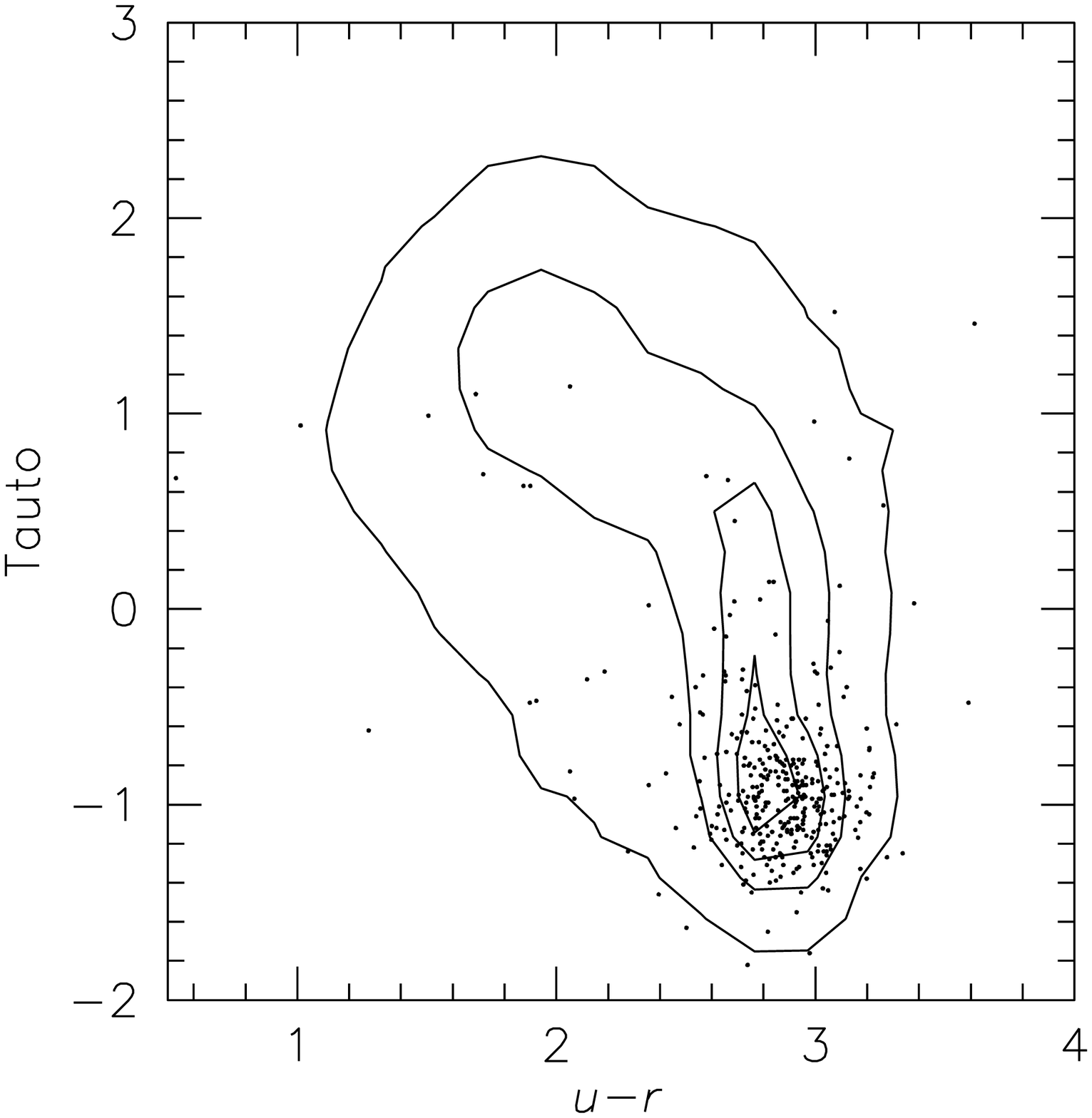}
\includegraphics[scale=0.2]{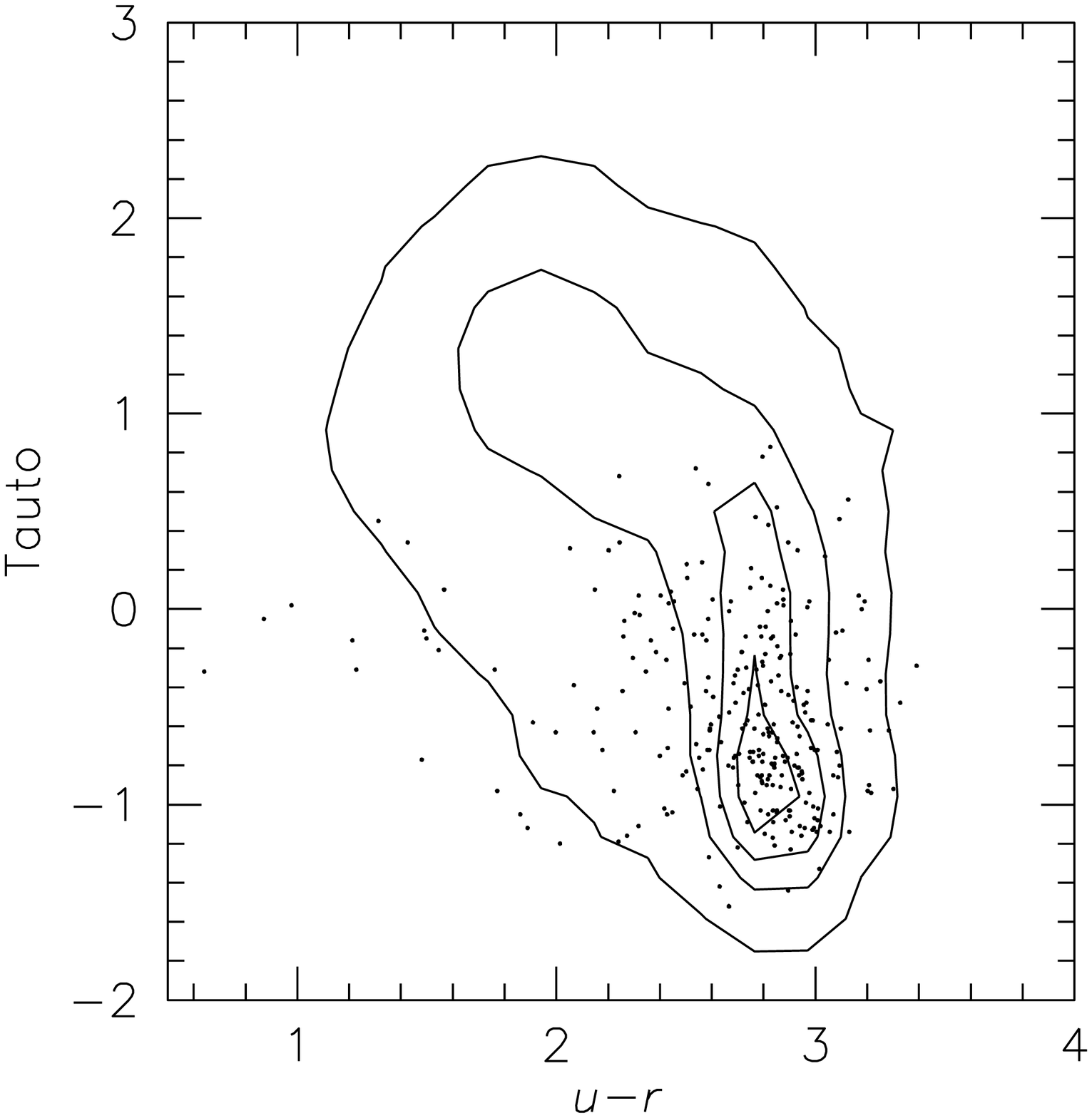}}
\centerline{\includegraphics[scale=0.2]{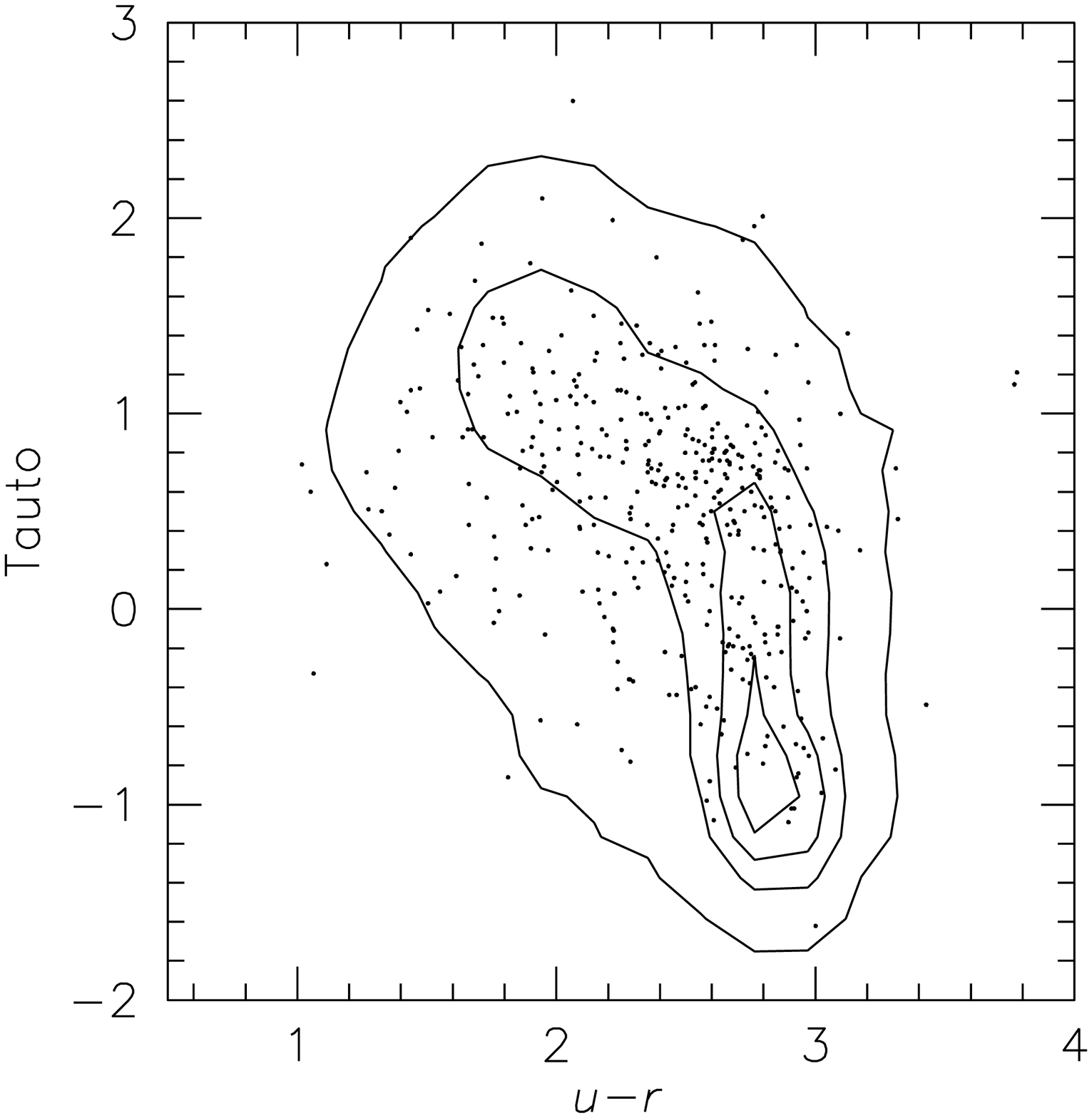}
\includegraphics[scale=0.2]{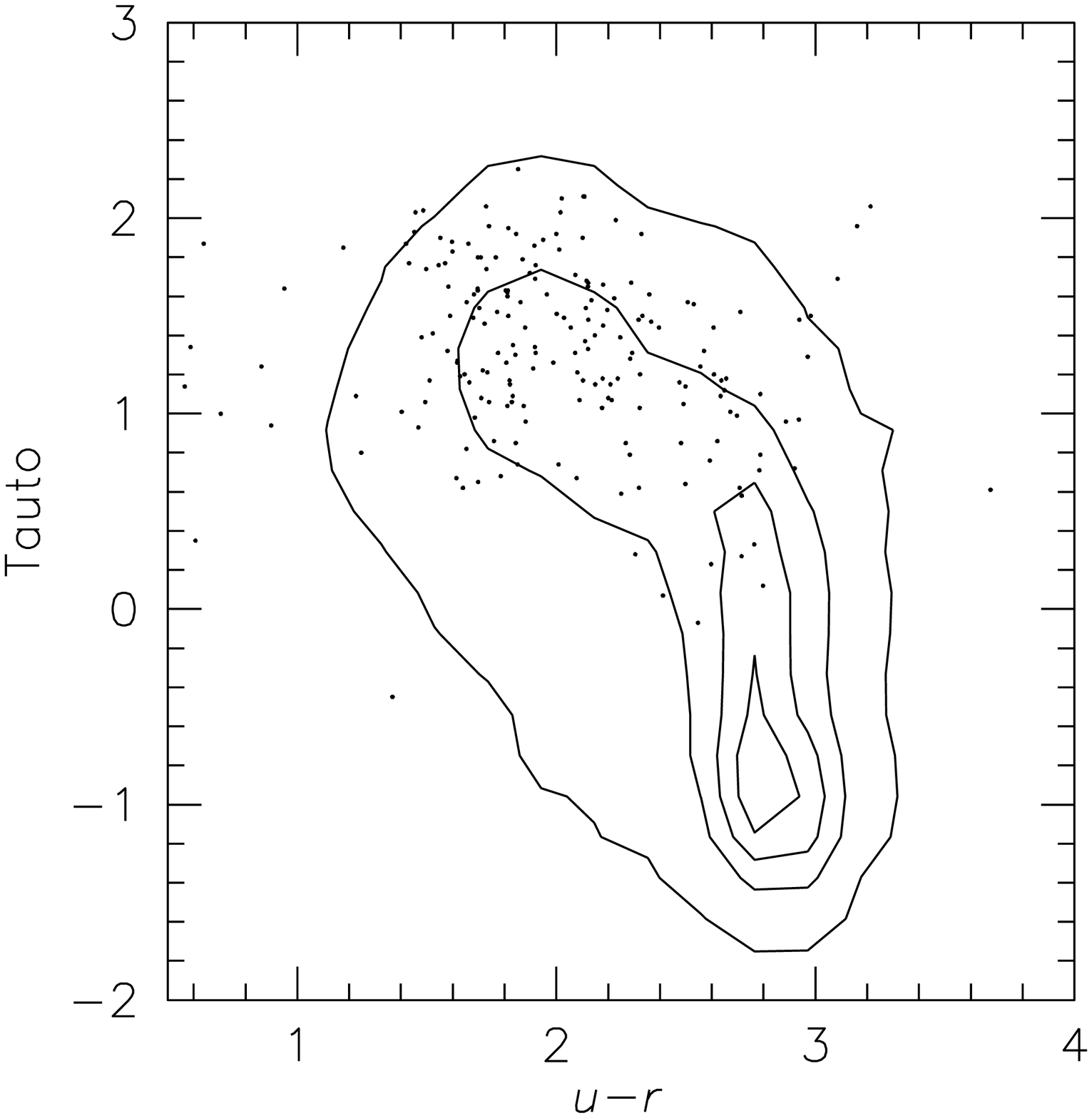}}
\caption{
\label{fig:tauto_ur}
 $Tauto$ is plotted against $u-r$. The extension of the distribution
 around $u-r$=2.8 is due to the inclination correction adopted in $Tauto$.
 The contours show distribution of all
 galaxies in the volume limited sample.  Points in each
 panel show the distribution of each morphological type of galaxies
 classified by eye (Shimasaku et al. 2001; Nakamura et
 al. 2003). Ellipticals are in the upper left panel. S0, Sa and Sc are
 in the upper right, lower left and lower right panel, respectively.   
}\end{figure}

\begin{figure}
\centerline{\includegraphics[scale=0.2]{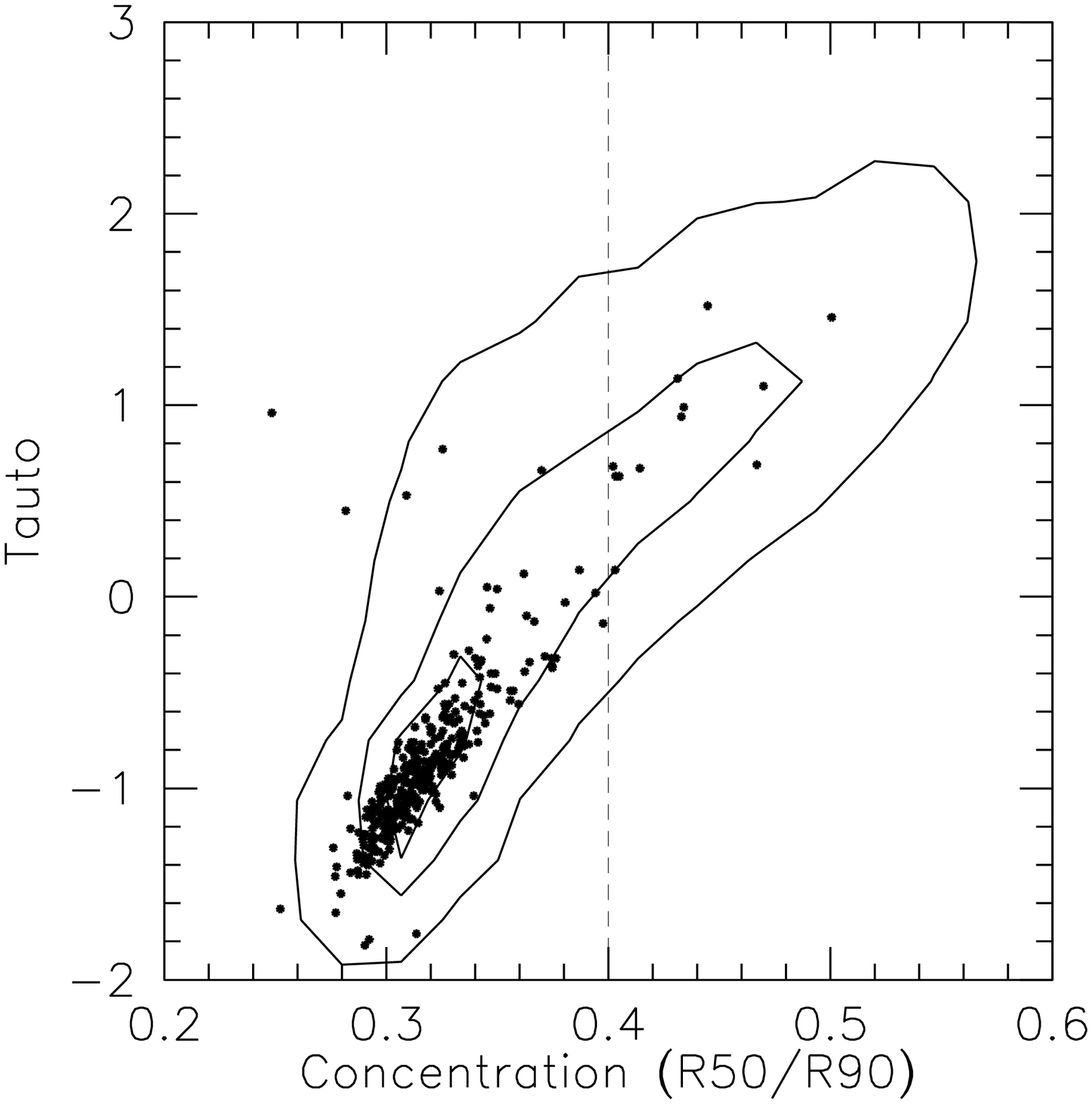}
\includegraphics[scale=0.2]{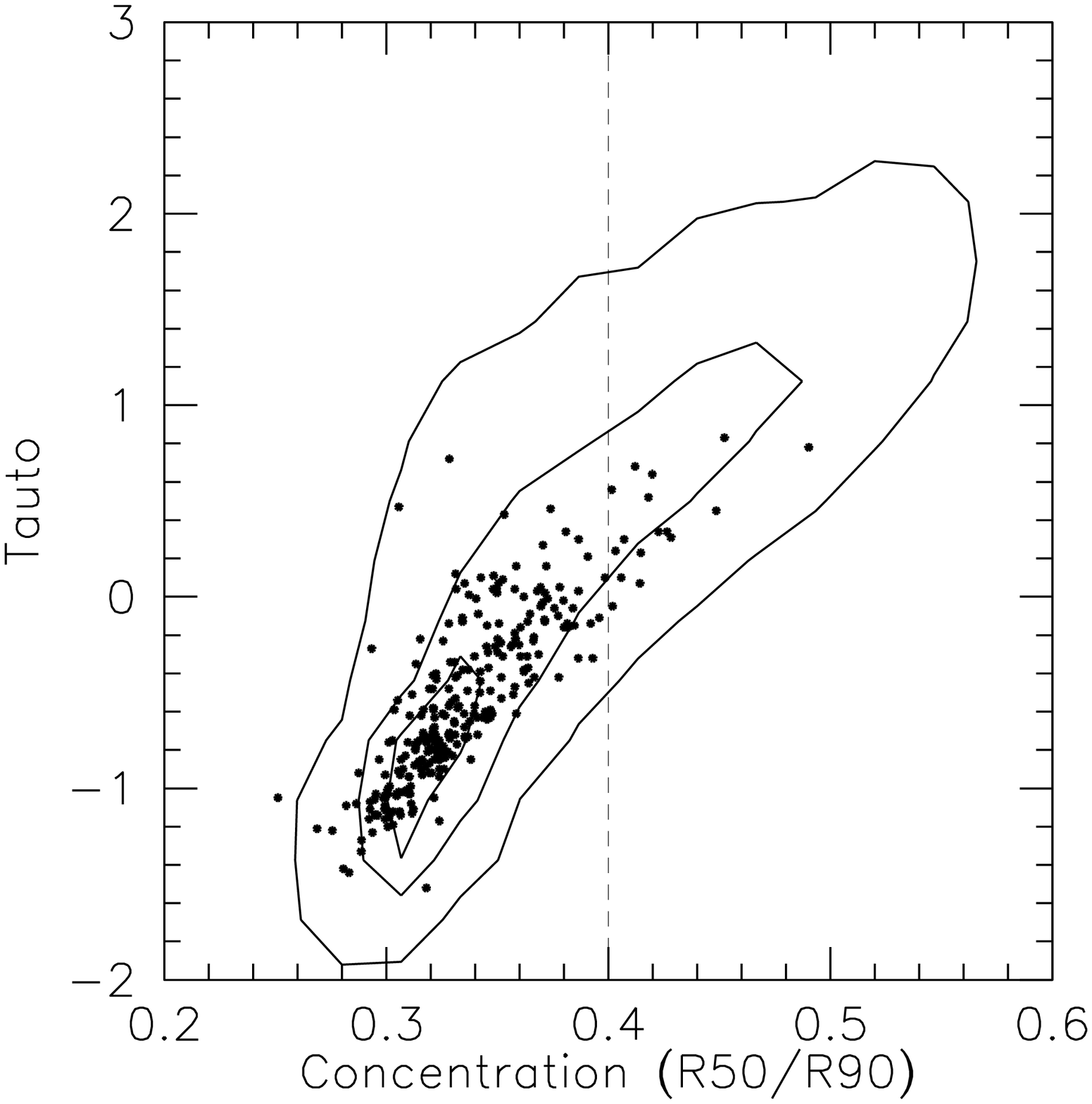}}
\centerline{\includegraphics[scale=0.2]{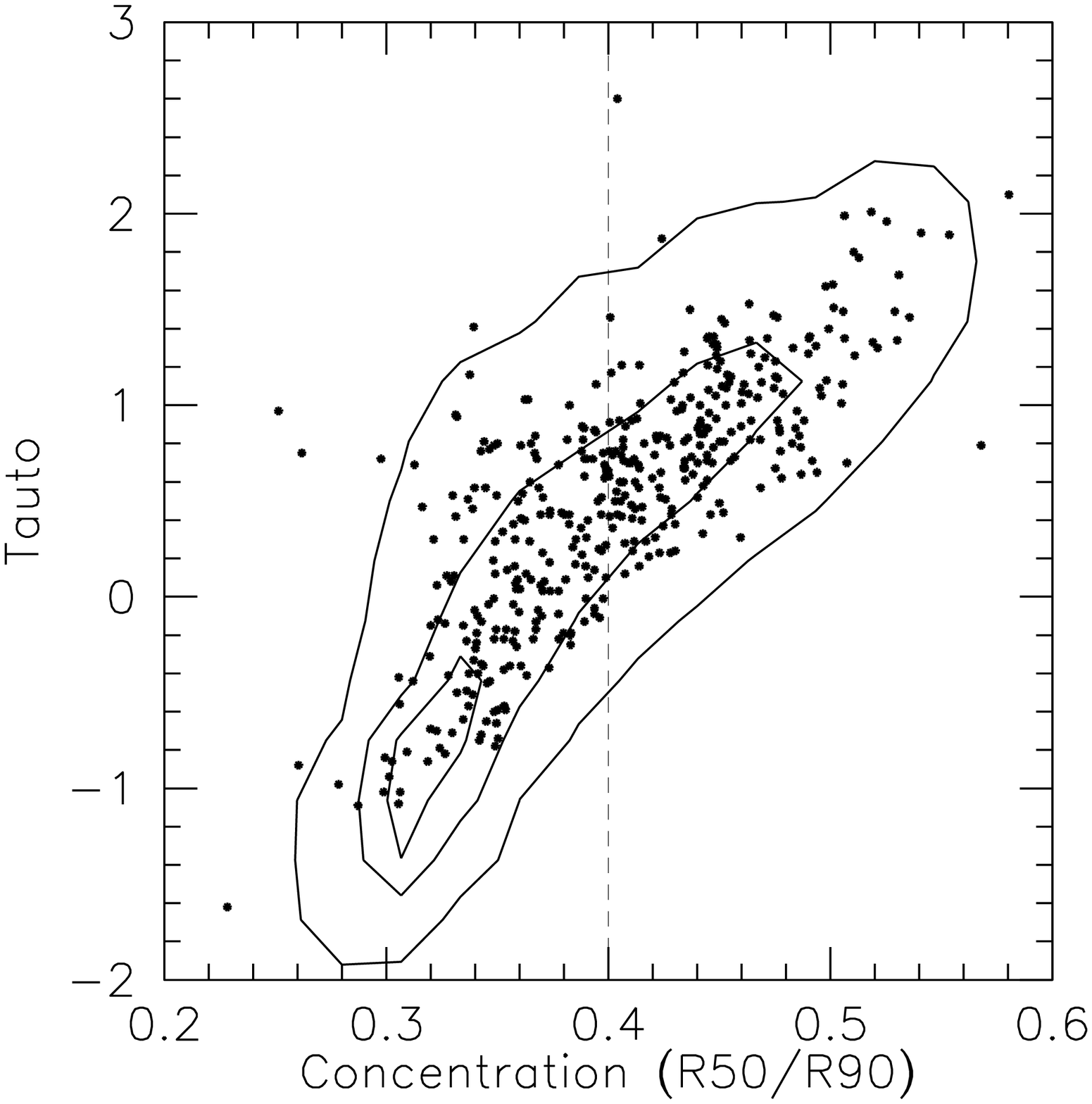}
\includegraphics[scale=0.2]{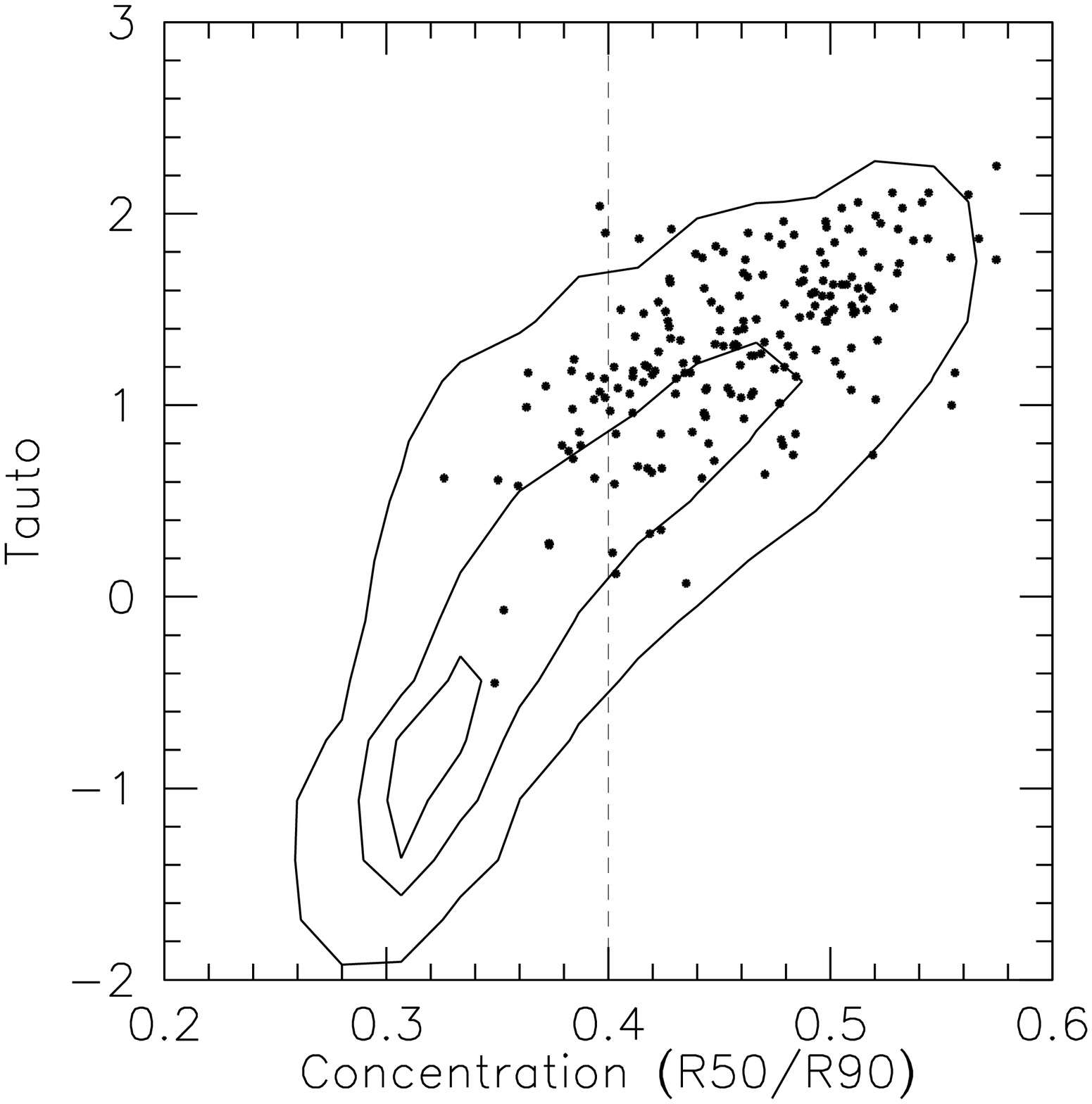}}
\caption{
 $Tauto$ is plotted against $Cin$. The contours show distribution of all
 galaxies in the volume limited sample. A good correlation between
 two parameters is seen. The extension of the distribution to the upper left
 panel is due to the inclination correction of $Tauto$. Points in each
 panel show the distribution of each morphological type of galaxies
 classified by eye (Shimasaku et al. 2001; Nakamura et
 al. 2003). Ellipticals are in the upper left panel. S0, Sa and Sc are
 in the upper right, lower left and lower right panel, respectively.   
}\label{fig:tauto_concent_each_type}
\end{figure}

\begin{figure}
\centerline{\includegraphics[scale=0.2]{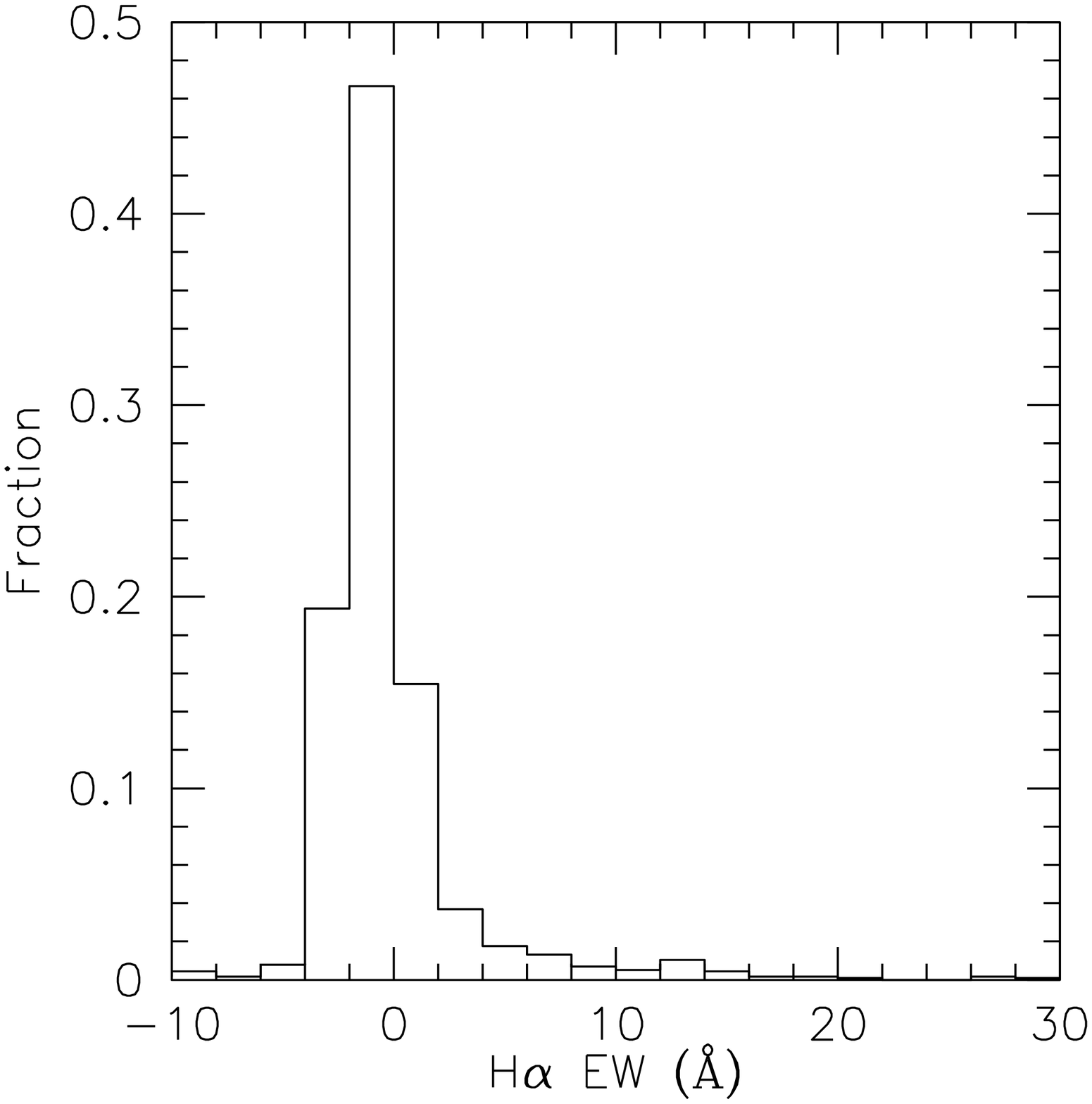}
\includegraphics[scale=0.2]{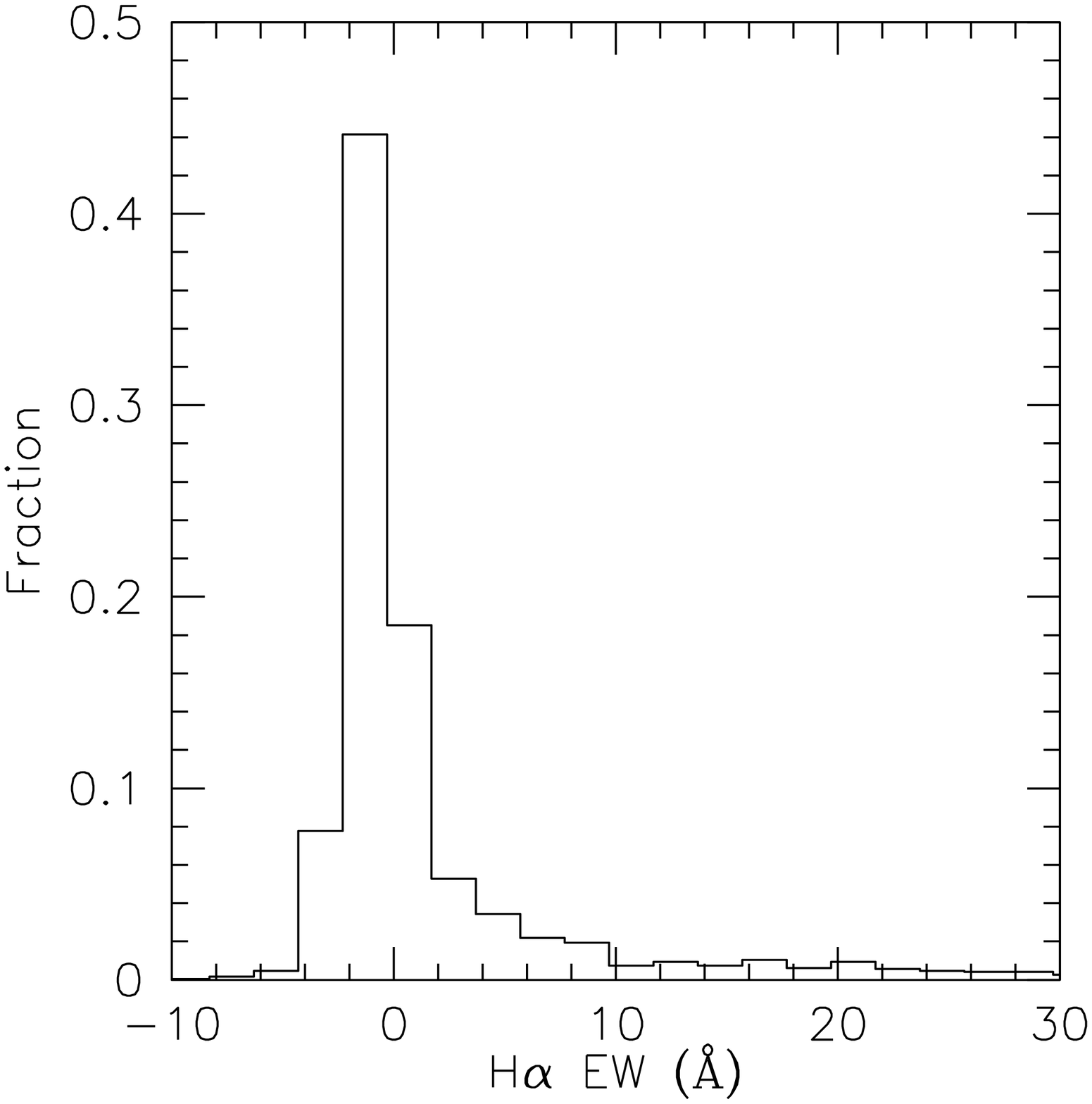}}
\centerline{\includegraphics[scale=0.2]{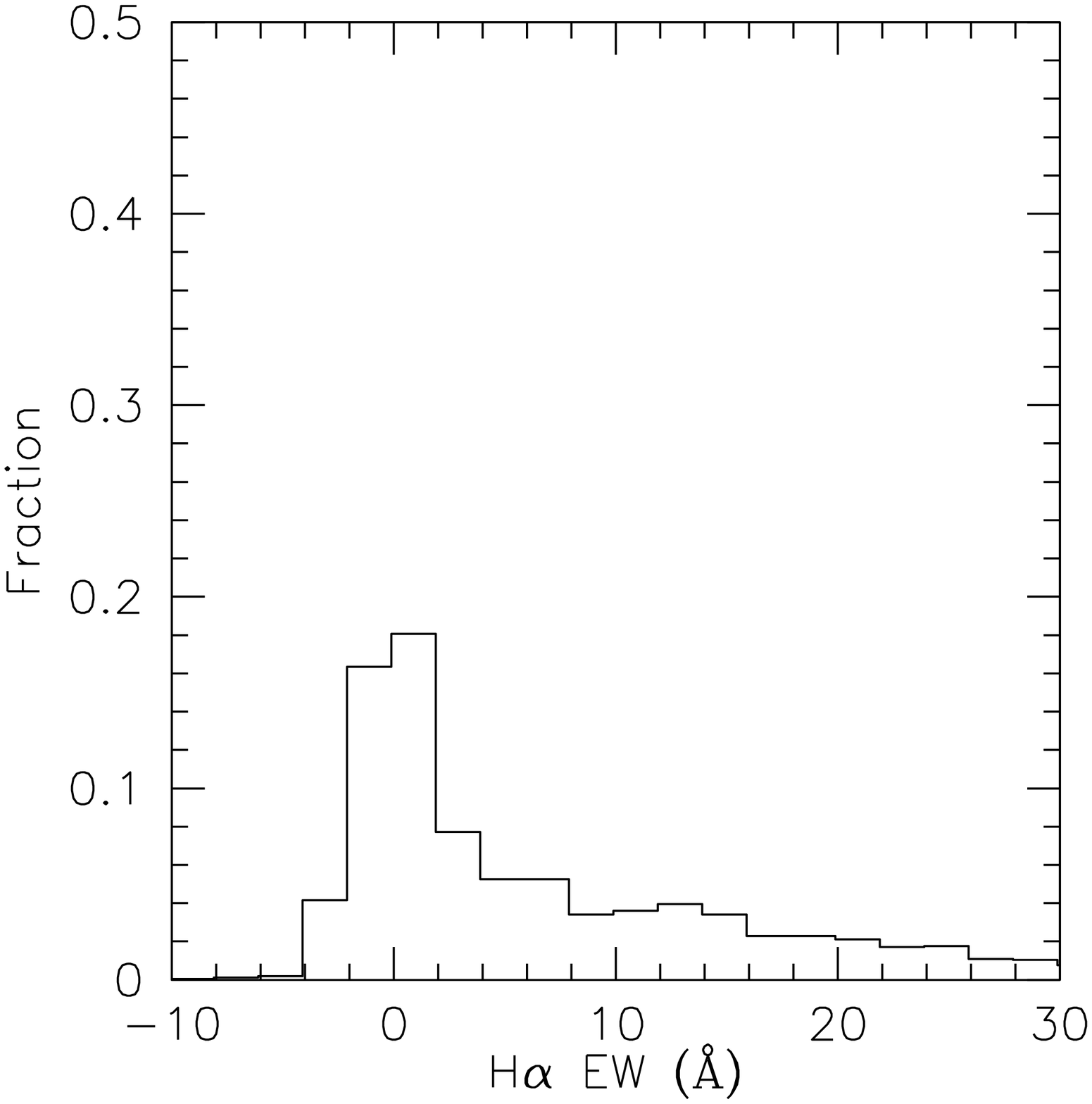}
\includegraphics[scale=0.2]{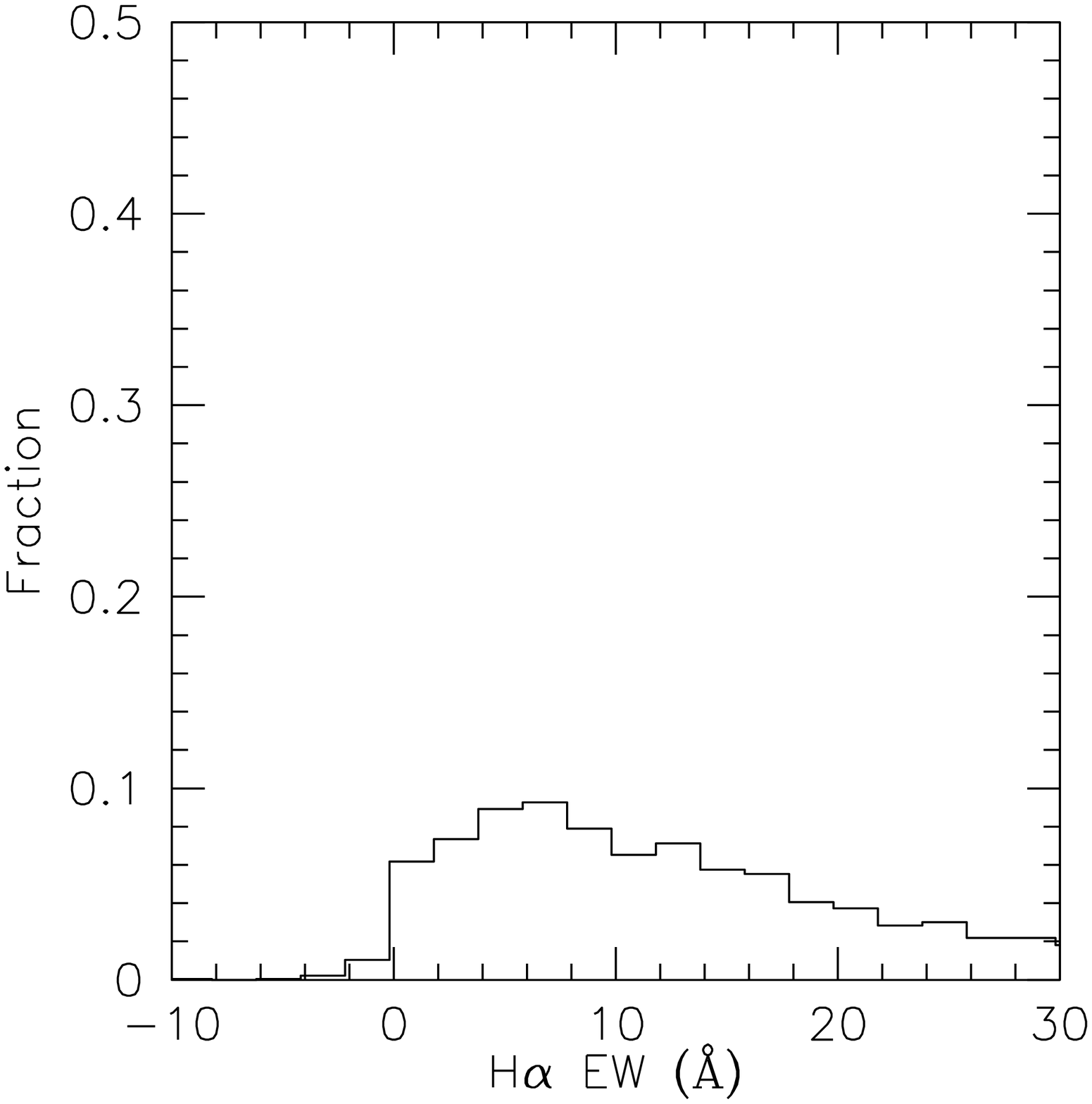}}
\caption{
\label{fig:ha}
 Distribution of H$\alpha$ EW for four types classified with
 $Tauto$. 
 A line in each
 panel shows the distribution of each morphological type of galaxies
 classified by $Tauto$. Early-type galaxies are in the upper left
 corner. Intermediate-type, early-disc and late-disc galaxies are
 in the upper right, lower left and lower right panel, respectively.  
 An increase of H$\alpha$ EW
 toward later type galaxies suggests that our morphological
 classification works well. 
}\end{figure}

 In figure \ref{fig:iskra}, $Cin$ is plotted against $u-r$ colour.
 Strateva et al. (2001) pointed out that $u-r$=2.2 serves as a good
 galaxy type classifier as well.  The
 distribution shows two peaks, one for elliptical galaxies at around
 ($u-r,Cin$)=(2.8,0.35), and one for spiral galaxies at around
 ($u-r,Cin$)=(2.0,0.45).  Our criterion at $Cin$=0.4 is located
 right between these peaks and separates these two populations
 well. 
 In figure  \ref{fig:tauto_ur},
 $Tauto$ is plotted against $u-r$ colour in four separate panels. 
 Due to the inclination
 correction of $Tauto$, two populations degenerated in $u-r$ colour (around $u-r$=2.8)
 are now separated into early-type and early-disc galaxies, which is
 one of the major improvements of $Tauto$ against $Cin$. Overplotted points are
 galaxies classified by eye (Shimasaku et al. 2001; Nakamura et
 al. 2003). The upper left, upper right, lower left and 
 lower right panels show elliptical,  S0, early-spiral and late-spiral
 galaxies classified by eye, respectively. Compared with $Tauto$
 criteria to separate the galaxies ($Tauto=-$0.8, 0.1 and 1.0), the
 figure suggests that our criteria separate galaxies reasonably well.  
  The effect of inclination correction of $Tauto$ can be also seen
 in figure \ref{fig:tauto_concent_each_type}, where $Tauto$ is plotted against
 $Cin$. In addition to the nice correlation between the two parameters,
 there are galaxies with high $Tauto$ and low $Cin$ values in the upper
 left of the figure. Most of these are edge-on galaxies correctly classified by $Tauto$
 due to its inclination correction. 
 In figure \ref{fig:ha}, we plot H$\alpha$\ EW for four types of
 galaxies classified with $Tauto$. Later type galaxies show higher
 H$\alpha$ EWs, suggesting our galaxy classification criteria work
 well (see Kennicutt 1998).

\subsection{Local Galaxy Density Measurements}
  
\begin{figure}
\includegraphics[scale=0.4]{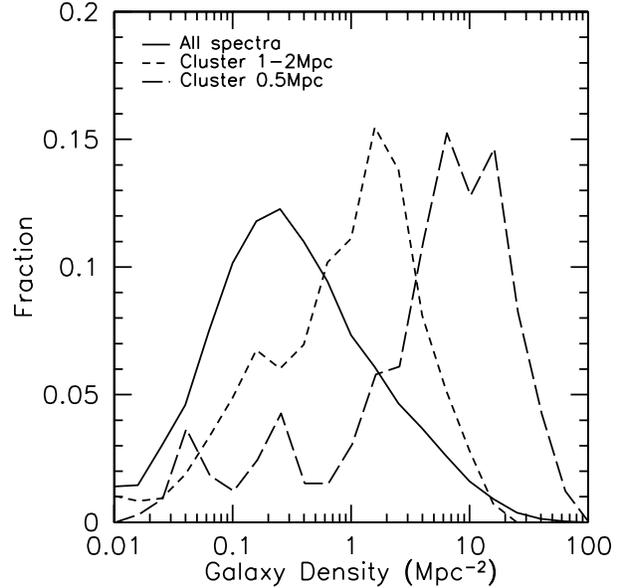}
\caption{
\label{fig:density_distribution} 
 Distribution of local galaxy density. The solid, dashed and dotted lines
 show distributions for all galaxies, galaxies within 0.5 Mpc from the nearest
 cluster and  galaxies between 1 and 2 Mpc from the nearest cluster, respectively.
}
\end{figure}

 We measure local galaxy density in the following way. 
 For each galaxy, we measure a projected distance to the 5th nearest
 galaxy within $\pm$1000 km s$^{-1}$ in redshift space among the volume limited sample
 (0.05$<z<$0.1, $Mr^*<-$20.5). 
 The 
 criterion for redshift space ($\pm$1000 km s$^{-1}$) is set to be generous to
 avoid galaxies with a large peculiar velocity slipping out of the
 density measurement, in other words, not to underestimate the density
 of cluster cores. Then, the number of galaxies ($N=5$) within the distance
 is divided by 
 the circular surface area with the radius of the distance to the 5th
 nearest galaxy. 
 When the projected area touches
 the boundary of the data, we corrected the 
 density by correcting the area
 to divide. Since we have redshift information for all of the sample
 galaxies, our density measurement is a
 pseudo-three dimensional density measurement and free from the
 uncertainty in background subtraction.  
 In figure \ref{fig:density_distribution}, we present distributions of
 this local galaxy density for all galaxies, galaxies within 0.5 Mpc
 from a cluster centre and galaxies between 1 and 2 Mpc from a cluster centre. In
 measuring distance form a cluster, we use the C4 cluster catalog
 (Miller et al. 2003). Part of the catalog is also presented in Gomez et
 al. (2003). For each galaxy, the distance from the nearest cluster
 centre is measured on a 
 projected sky for galaxies within $\pm$1000 km s$^{-1}$ from a cluster redshift.

\section{Results}\label{results}
\subsection{The Morphology Density Relation in the SDSS Data}\label{morphology_density}
  
\begin{figure}
\begin{center}
\includegraphics[scale=0.4]{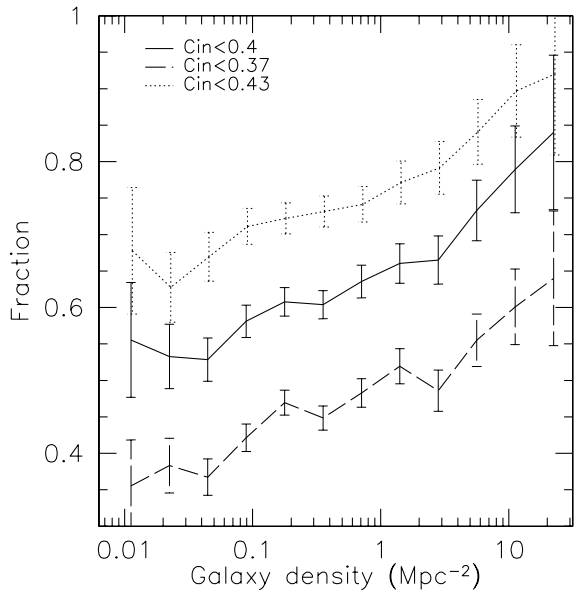}
\end{center}\caption{
\label{fig:ann_cin}
 The morphology-density relation for three criteria of $Cin$. Fractions of
 early-type galaxies are plotted against local galaxy density. Three
 criteria are $Cin<$0.4, $Cin<$0.43 and $Cin<$0.37 in the solid, dashed and
 dotted lines, respectively. We do not include galaxy density $>$30
 Mpc$^{-2}$ in this plot since above this density there are only a few
 galaxies in each bin.
}
\end{figure}

 In figure \ref{fig:ann_cin}, we use $Cin$ to present the ratio of the
 number of early type galaxies to that of all galaxies as a
 function of the local galaxy density. The solid line shows the
 ratio of early type galaxies using $Cin$=0.4 as a separator,
 which separate elliptical galaxies and spiral galaxies well as shown in
 figure \ref{fig:iskra}. 
 The fraction of
 early type galaxies clearly increases with increasing density. In the
 least dense region, only 55\% are early type, whereas in the densest
 region, almost 85\% are early type galaxies. Furthermore, it is
 interesting to see 
 that around galaxy density 3 Mpc$^{-2}$, the slope of the
 morphology-density relation abruptly becomes flatter.  3 Mpc$^{-2}$ is
 a characteristic density for cluster perimeter (1-2 Mpc; see figure
 \ref{fig:density_distribution}) and also coincides with the density
 where star formation rate (SFR) of galaxies changes as studied by Lewis et
 al. (2002) and Gomez et al. (2003). 
  To see the dependence of the relation to the
 choice of our morphological criterion ($Cin$=0.4), we use slightly different criteria for dashed
 and dotted lines, which use $Cin$=0.37 and $Cin$=0.43 as a criterion, respectively. 
 The former criterion is a little biased to spiral galaxies and the
 latter to elliptical galaxies. As is seen in the figure, both dotted
 and dashed lines show the morphology-density relation, but in somewhat
 flatter way than the solid line, indicating the effect of the
 contamination from spiral galaxies in case of $Cin$=0.43
 (incompleteness in case of $Cin$=0.37). In the density below 3
 Mpc$^{-2}$, the steepness of three slopes are almost identical. However, at the densest
 region, only 65\% is early-type galaxies in case of $Cin$=0.37,
 whereas almost 90\% is early-type galaxies in case of  $Cin$=0.43.
 Thus, the absolute amount of early-type galaxies are strong function of
 $Cin$ criterion. Therefore, careful attention to $Cin$ criterion is
 needed when comparing to other work such as computer simulations and
 other observational data.

\begin{figure}
\begin{center}
\includegraphics[scale=0.4]{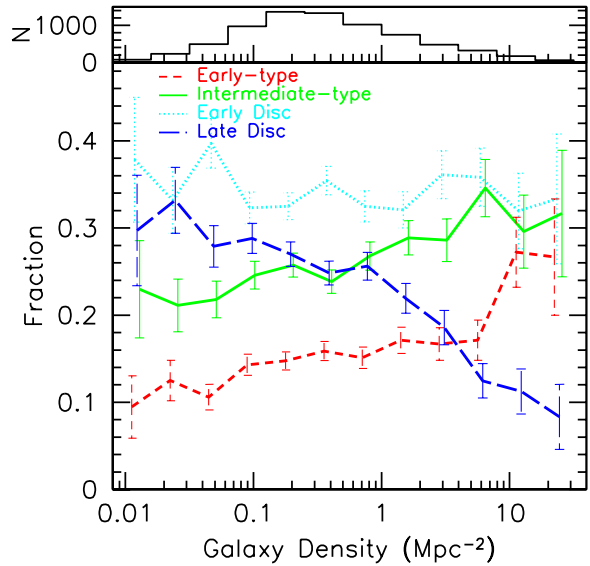}
\end{center}
\caption{
\label{fig:md_ann_ytype}
 The morphology-density relation for four types of galaxies classified
 with $Tauto$.  
 The short-dashed, solid,  dotted and long-dashed lines represent 
  early-type,  intermediate-type, early-disc and late-disc galaxies,
 respectively.
 The histogram in the upper panel shows the numbers of galaxies in each
 bin of local galaxy density.
}
\end{figure}

\begin{figure}
\begin{center}
\includegraphics[scale=0.4]{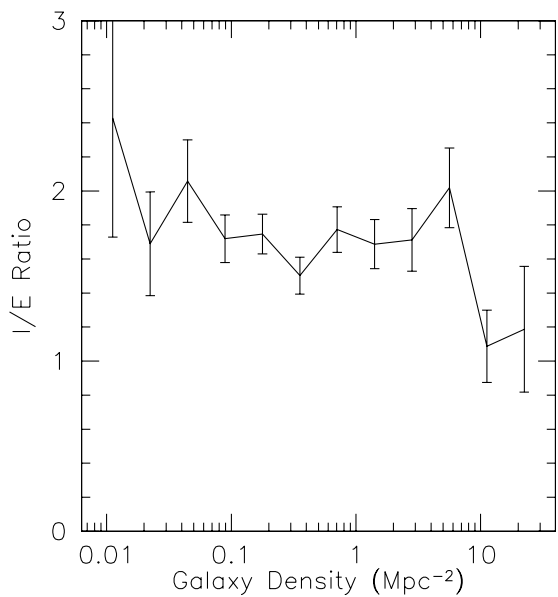}
\end{center}
\caption{
\label{fig:md_es0}
 Intermediate-type to early-type number ratio as a function of local galaxy density.
}
\end{figure}

 In figure \ref{fig:md_ann_ytype}, the ratio of four morphological
 types of galaxies are plotted against galaxy density using
 $Tauto$. The short-dashed, solid, dotted and long-dashed lines represent
 early-type ($E$), intermediate ($I$), early-disc ($ED$) and late-disc ($LD$)
  galaxies, respectively.  The error bars are calculated using Poisson
  statistics.  The histogram in the upper panel shows the numbers of
  galaxies in each bin of the local galaxy density. 
 The decline of
 late-disc fraction in going toward high density is seen. Early-disc
 fractions stays almost constant. 
 Intermediate-type fractions dramatically increase toward higher density, but
 declines somewhat at the two highest density bins. Early-type fractions
 show mild 
 increase with increasing density and radically increase at the two highest
 density bins.  
 In the figure, there exist two characteristic densities where the
 relation radically changes. Around galaxy density 1 Mpc $^{-2}$,
 corresponding to the cluster infalling region
 (Fig. \ref{fig:density_distribution}), the slope for late-disc
 suddenly goes down and  the slope for intermediate-type rises up.
 At around galaxy density 6 Mpc $^{-2}$, corresponding to the
 cluster core region (Fig. \ref{fig:density_distribution}),
 intermediate-type fractions 
 suddenly goes down and early-type fractions show a sudden increase. 
 To clarify this second change in early-type and intermediate-type fractions, we plot
 intermediate-type to early-type number ratio against local galaxy density in figure
 \ref{fig:md_es0}. As is seen in the previous figure, $I/E$ ratio
 mildly increases from galaxy density 1 to 5 Mpc $^{-2}$, then suddenly
 declines after 6 Mpc $^{-2}$. We discuss the interpretation of this
 result in Section \ref{discussion}. The current results are derived
 from only $\sim$5\% of the final SDSS data.  When the SDSS is
 completed, the errors on each data point  will be reduced by approximately 80\%. 

 %

\subsection{Morphology-Radius Relation in the SDSS Data}
\label{Oct 30 18:22:01 2002}
\begin{figure}
\includegraphics[scale=0.4]{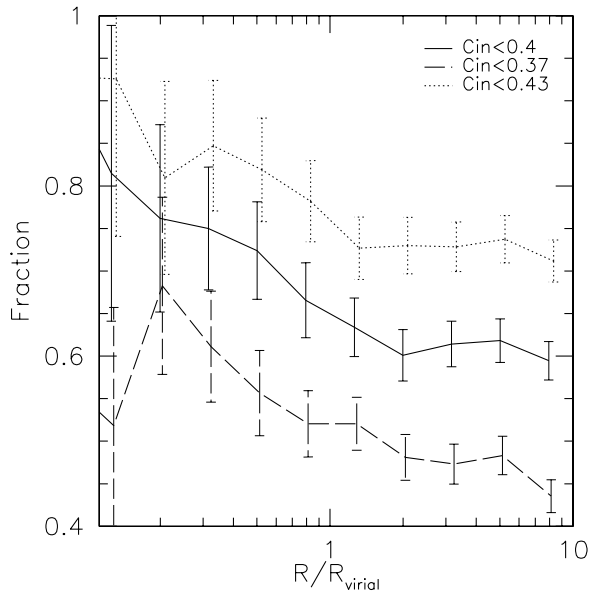}
\caption{
 The morphology-radius relation based on $Cin$. Fractions of early-type galaxies are plotted against cluster-centric radius to the nearest cluster.  
 Morphological criteria are $Cin<$0.4, $Cin<$0.43 and $Cin<$0.37 in the solid, dashed and dotted lines, respectively.
}\label{fig:mr_cin} 
\end{figure}

  In figure \ref{fig:mr_cin}, we plot early-type fractions classified with
  $Cin$, against a
  radius from the cluster centre. We use the C4 galaxy cluster catalog
  (Miller et al. 2003) when measuring the distance between a galaxy and
  the nearest cluster. 
  For each cluster, distances are measured by converting angular
  separation into physical distance for galaxies within
  $\pm$1000 km s$^{-1}$ in redshift space. For each galaxy, the distance to the
  nearest cluster is adopted as a distance to a cluster. The distance is
  converted to in a unit of one virial radius using velocity dispersion
  given in Miller et al. (2003) and  the equation given in Girardi et
  al. (1998). 
  The  morphological fraction for each radius bin is measured in the same way
  as the last section; the solid, dashed and dotted lines represents
  different criteria, $Cin=$0.4, 0.37 and 0.43, respectively. As seen in
  figure \ref{fig:iskra},   $Cin=$0.4 best separates elliptical and
  spiral galaxies. 
  In the figure, fraction of early-type galaxies
  decreases toward larger distance from a cluster centre. The relation
 becomes consistent with flat after 1 virial radius. 
  As in the case in figure \ref{fig:ann_cin}, three criteria show
  similar slope. However, absolute amount of early-type galaxies is a
  strong function of $Cin$ criteria. In case of $Cin$=0.4, early-type
  fractions increase from 60\% to almost 90\% toward a cluster centre.

\begin{figure}
\includegraphics[scale=0.4]{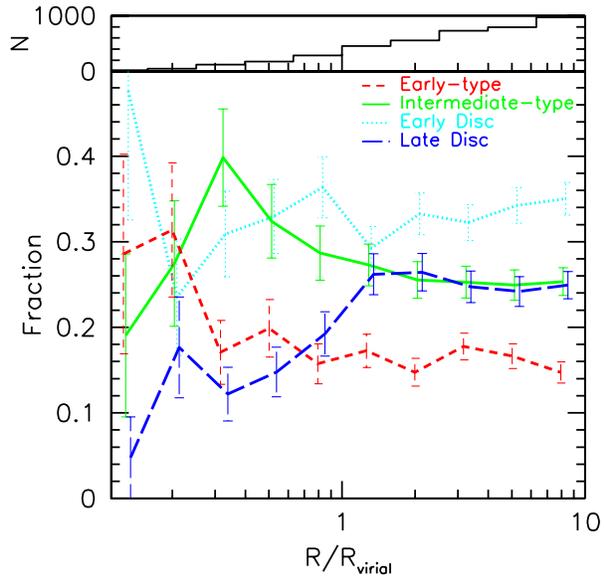}
\caption{
\label{fig:mr} 
 The morphology-radius relation based on $Tauto$. Fractions of each type
 of a galaxy is plotted against cluster-centric radius to the nearest
 cluster. The short-dashed, solid, 
  dotted and long-dashed lines represent 
 early-type,  intermediate-type, early-disc and late-disc galaxies,
 respectively.
  The histogram in the upper panel shows the numbers of galaxies in each
 bin of cluster-centric radius.
}
\end{figure}

\begin{figure}
\includegraphics[scale=0.4]{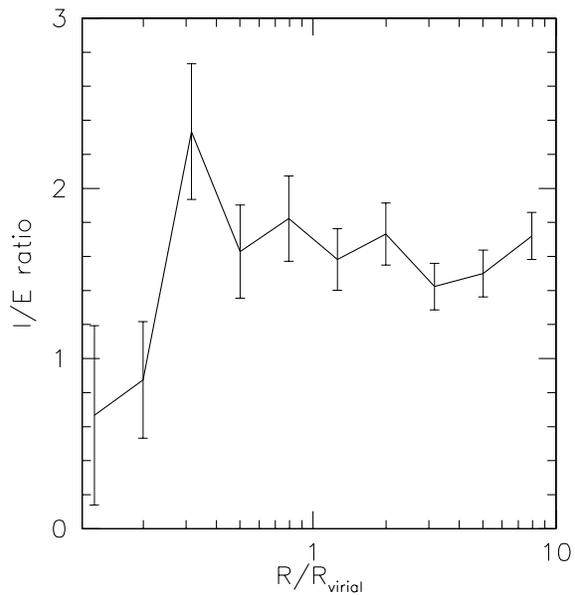}
\caption{
\label{fig:mr_es0} 
 Intermediate-type to early-type number ratio is plotted against cluster centric
 radius. The ratio decreases at the cluster core region.
}
\end{figure}

  In figure \ref{fig:mr}, we plot the morphology--cluster-centric-radius
  relation for four types of galaxies classified using
  $Tauto$. As is in figure \ref{fig:md_ann_ytype}, the short-dashed, solid,
  dotted and long-dashed lines represent 
  early-type,  intermediate-type, early-disc and late-disc galaxies,
  respectively.  The error bars are calculated using Poisson
  statistics.  The histogram in the upper panel shows the numbers of
  galaxies in each bin of cluster-centric radius.  
  Fractions of late-disc galaxies decrease toward smaller radius,
  whereas fractions of early-type and intermediate galaxies increase toward a
  cluster centre. In the figure, three characteristic radii are
  found. Above 1 virial radii, four lines are consistent with flat,
  suggesting that physical mechanisms responsible morphological change do not
  work beyond this radius. Between 0.3 and 1 virial radius, intermediate-type fractions
  mainly increases toward a cluster centre. Late- and early-disc galaxies
  show corresponding decrease. Interestingly intermediate-type fractions increase more
  than early-type fractions. Below 0.3 virial radius, early-type fractions
  dramatically increase and intermediate-type fractions decrease in turn. 
  To further clarify the change between intermediate-type and early-type fractions, we
  plot intermediate to early-type number ratio in figure \ref{fig:mr_es0}. As is
  seen in the figure \ref{fig:md_es0}, the ratio slightly increase between 1 and 0.3
  virial radius toward a cluster centre. At 0.3 virial radius, slope
  changes radically and the ratio decreases toward  a cluster centre. We
  interpret these findings in Section \ref{discussion}. 

\subsection{Density vs. Radius}\label{density_radius}
\begin{figure}
\centerline{\includegraphics[scale=0.18]{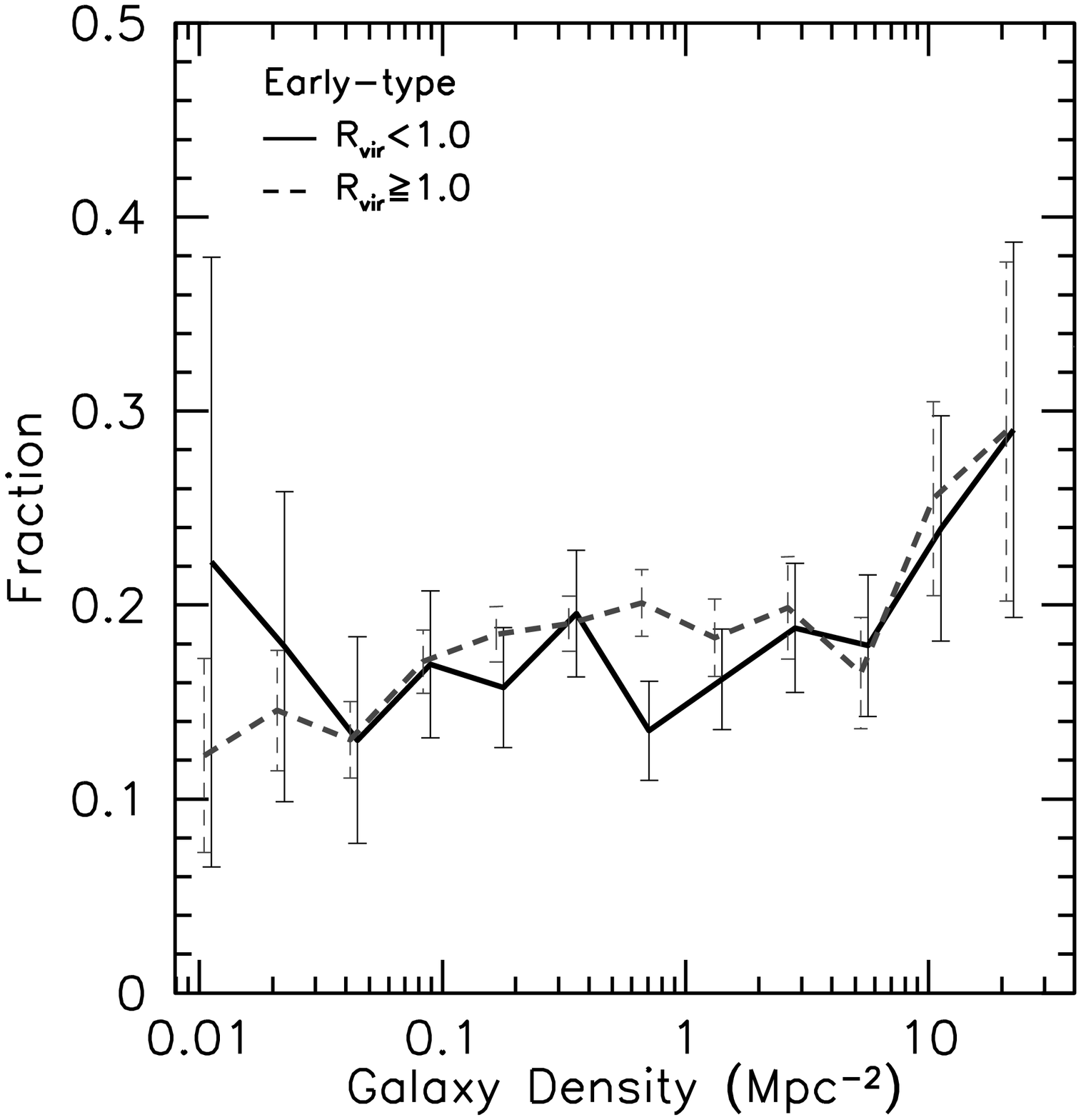}
\includegraphics[scale=0.18]{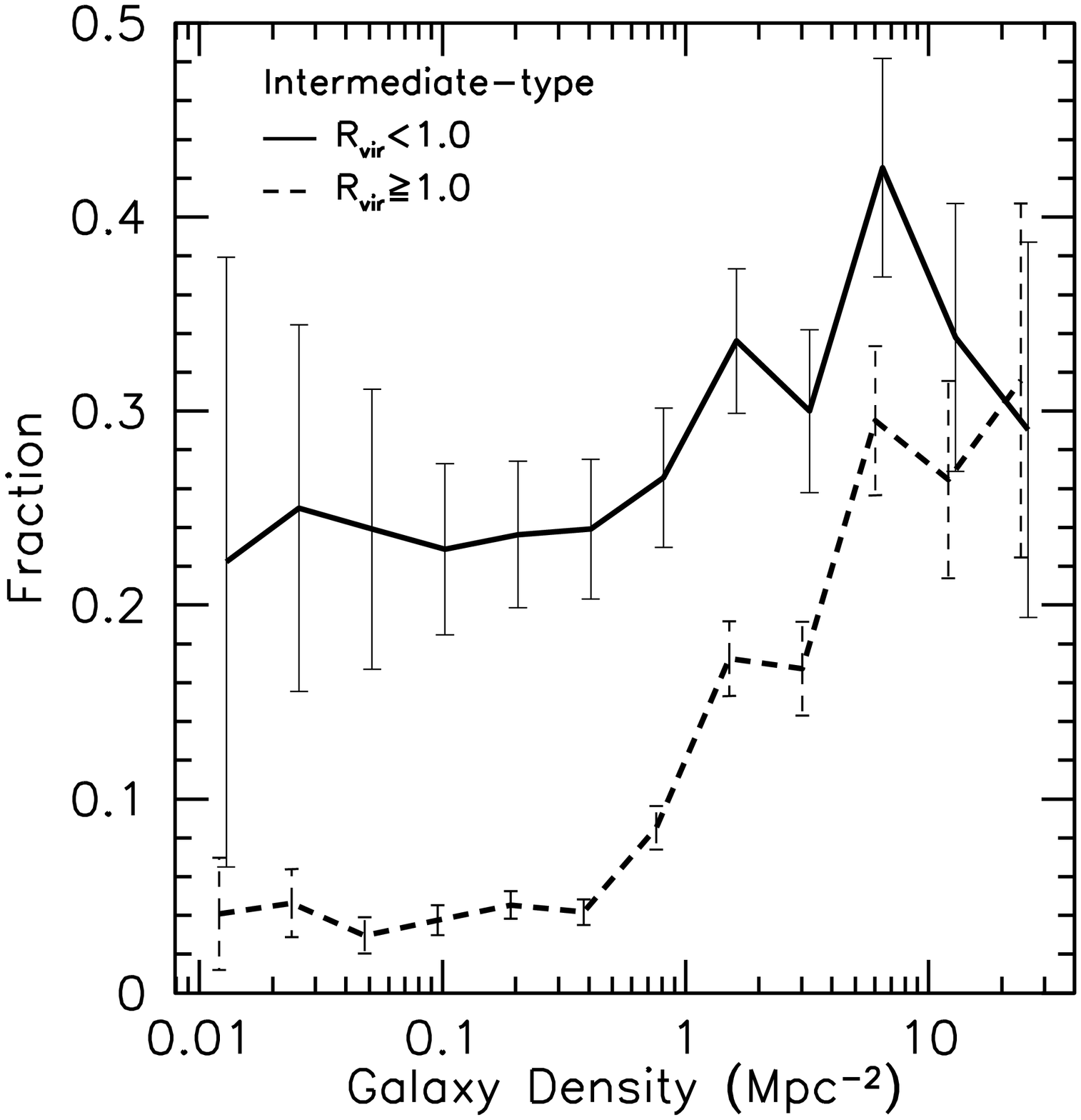}}
\centerline{\includegraphics[scale=0.18]{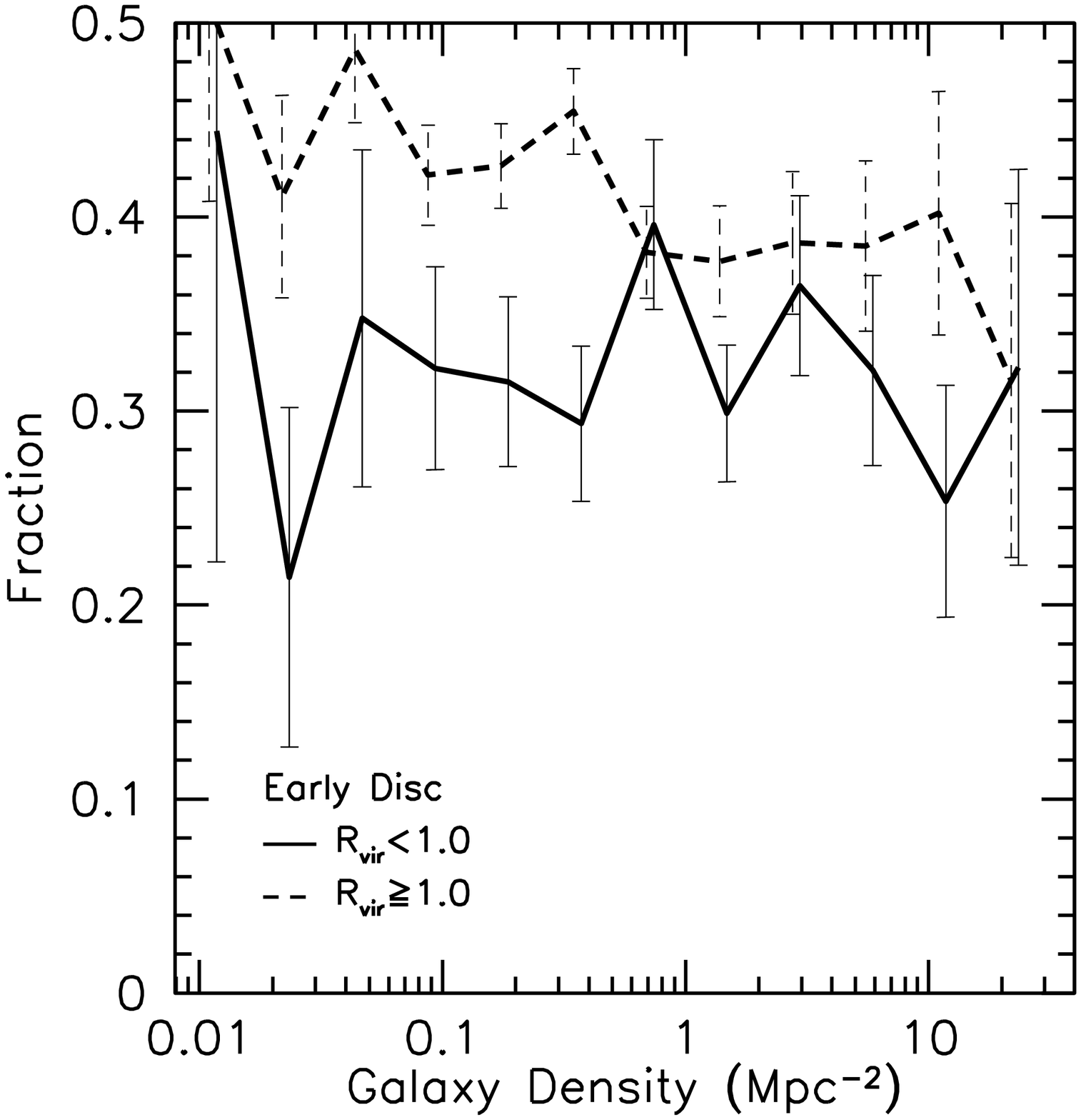}
\includegraphics[scale=0.18]{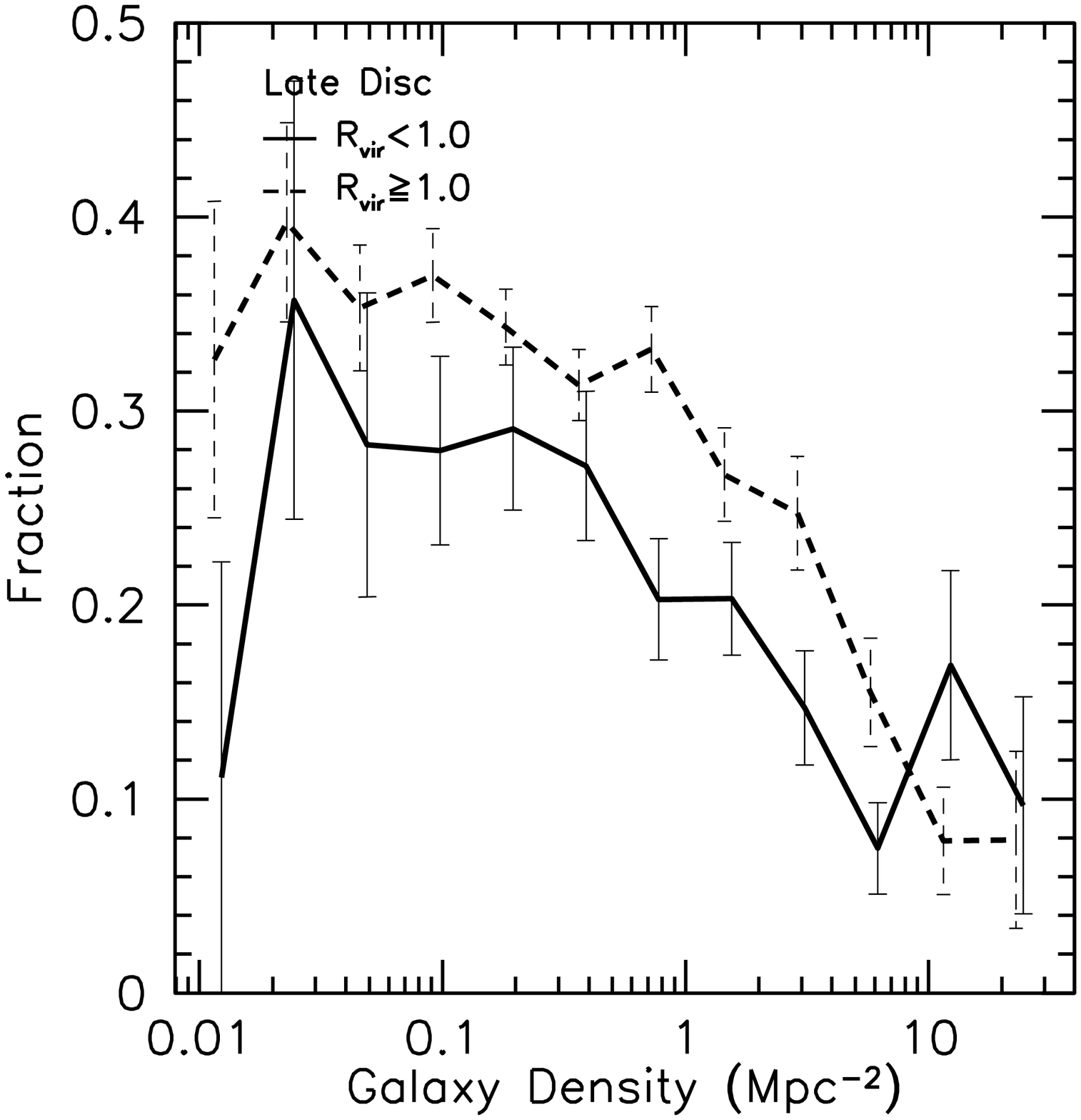}
}
\caption{\label{fig:md_rad}
 The morphology-density relation divided by $R_{vir}=1$.  The solid
 lines represent galaxies with $R_{vir}<1$. The dashed lines represent
 galaxies with  $R_{vir}\geq1$.
 Early-types
 are in the upper left panel. Intermediate-types, early discs and late discs are
 in the upper right, lower left and lower right panel, respectively. 
}
\end{figure}

\begin{figure}
\centerline{\includegraphics[scale=0.18]{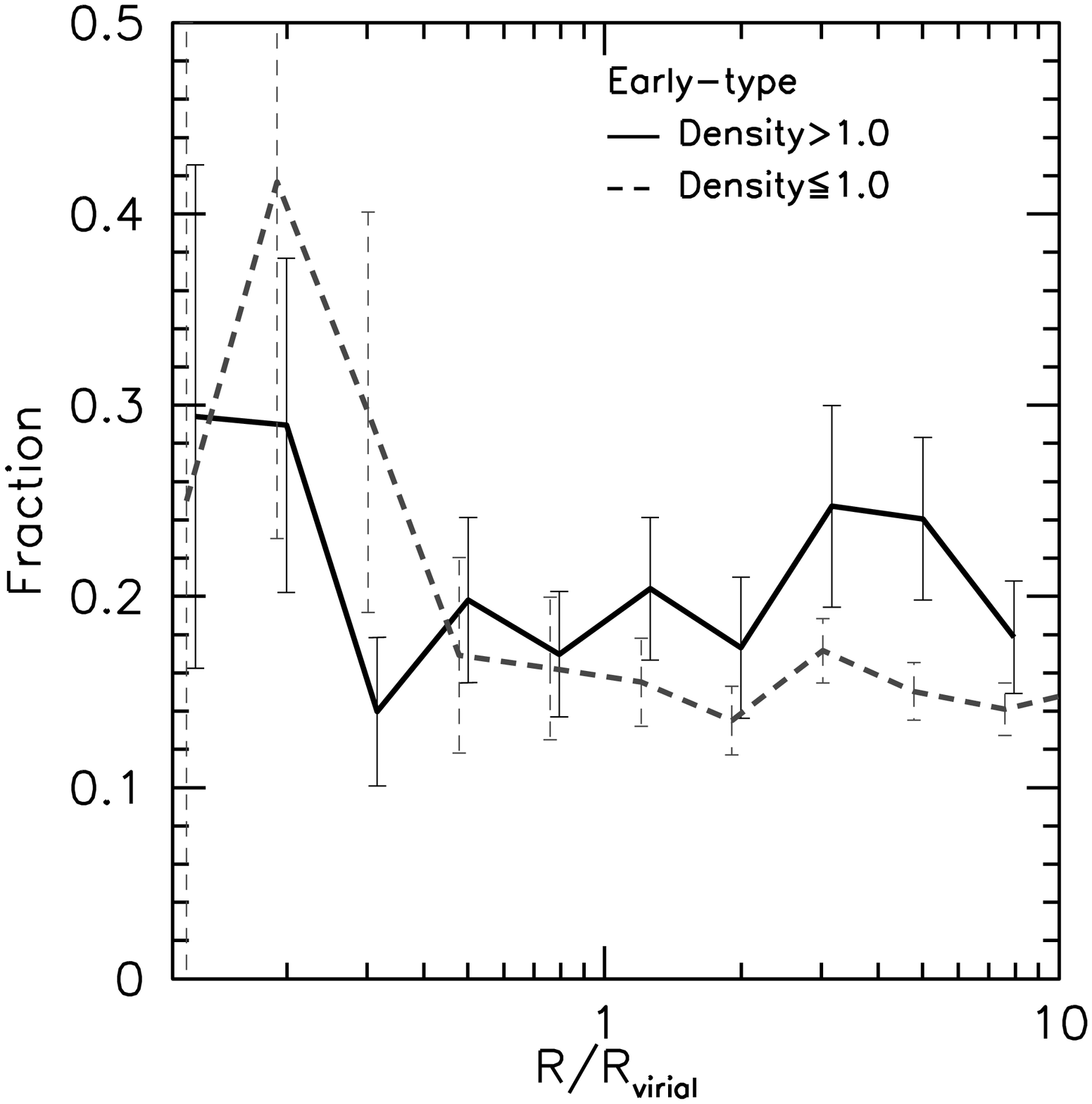}
\includegraphics[scale=0.18]{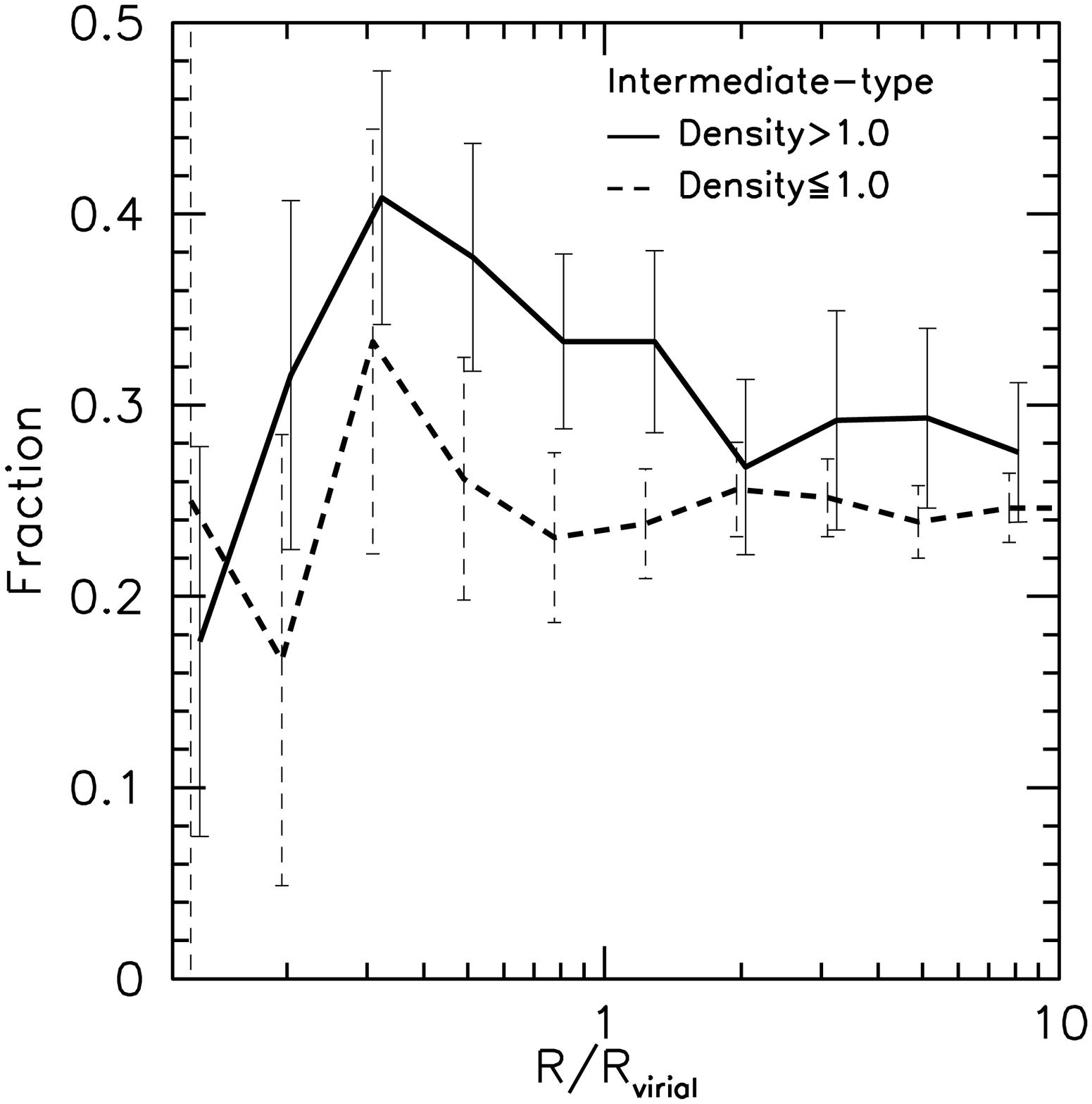}}
\centerline{\includegraphics[scale=0.18]{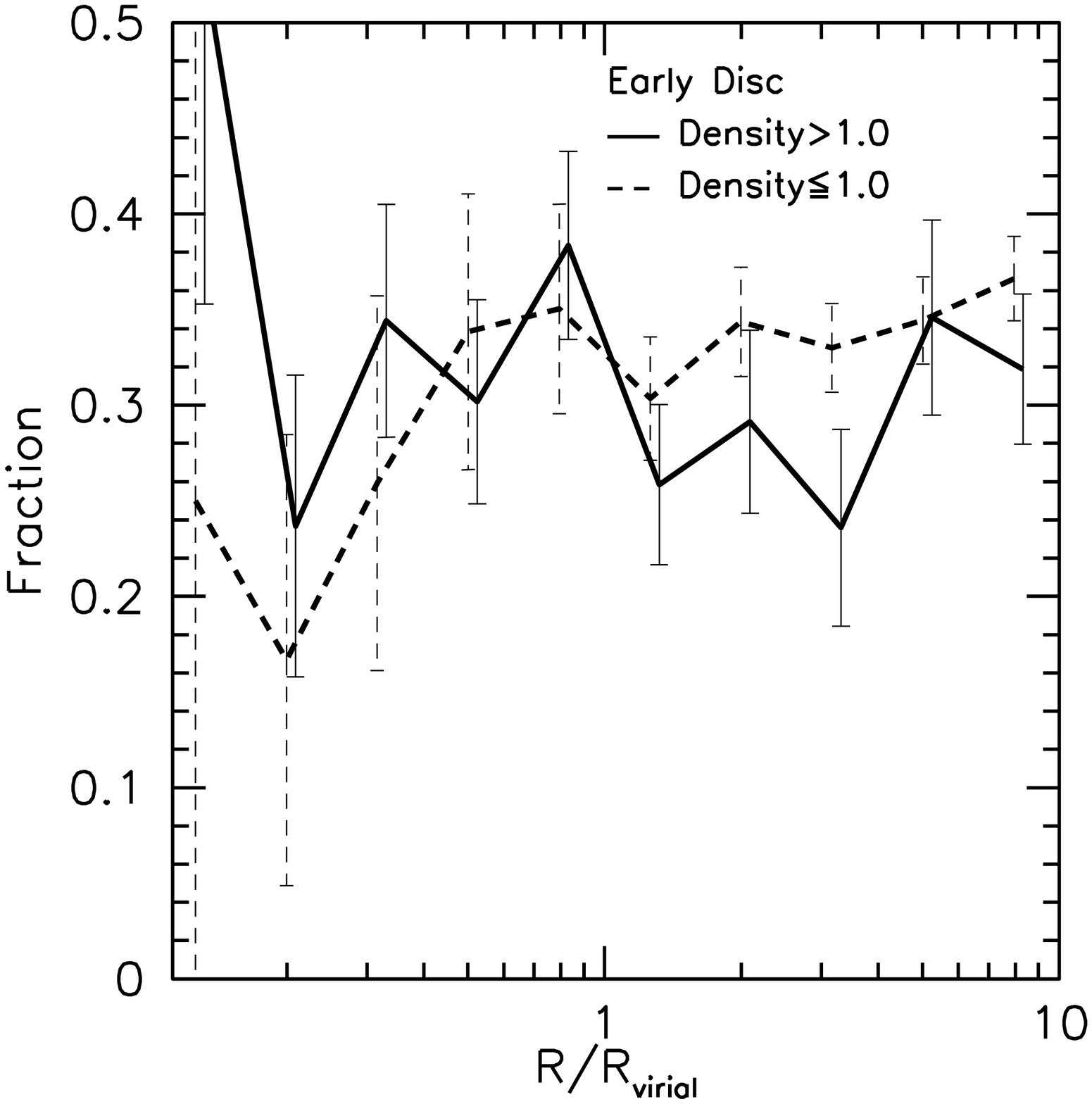}
\includegraphics[scale=0.18]{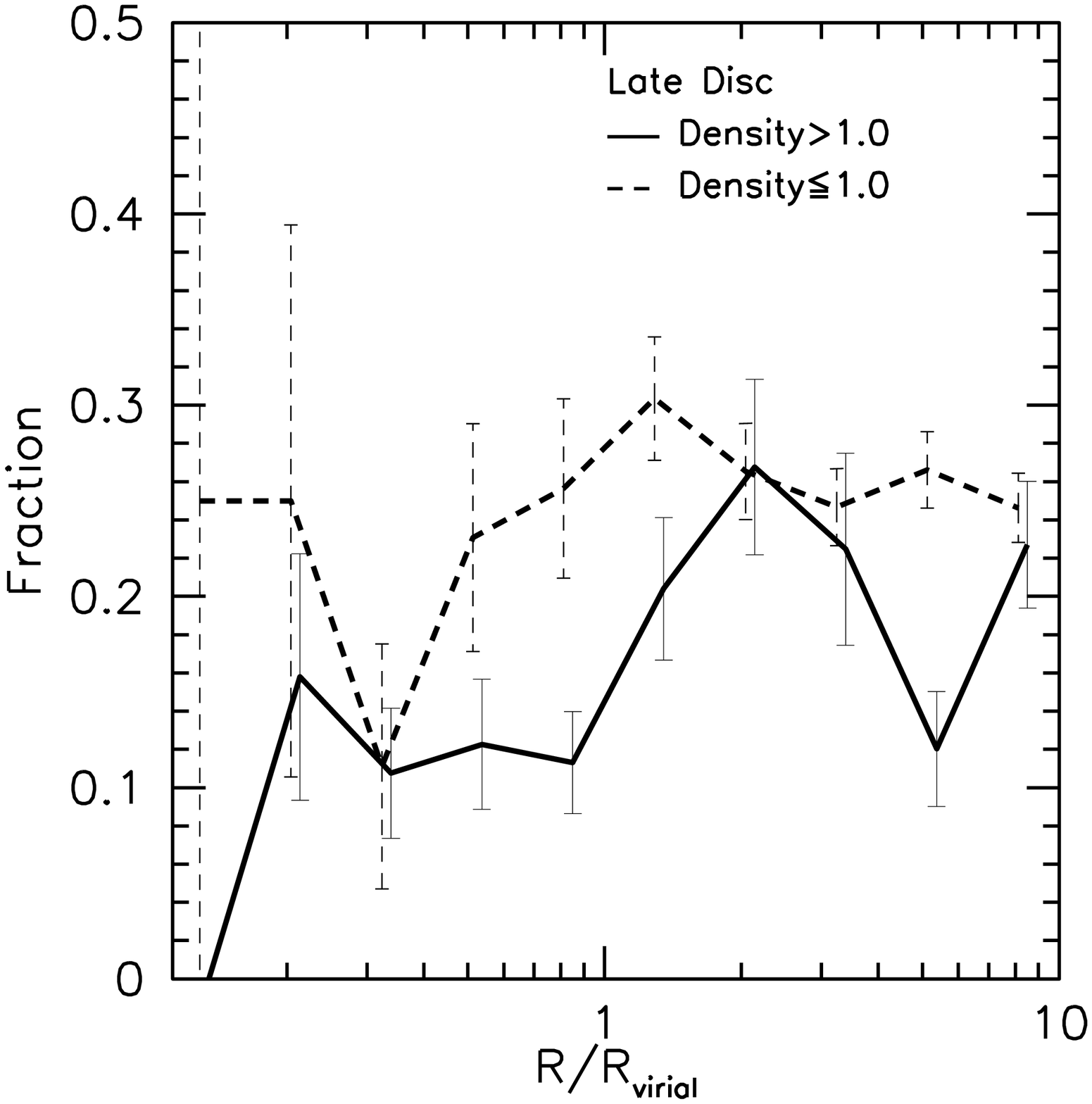}
}
\caption{\label{fig:mr_5th}
 The morphology--cluster-centric-radius relation divided at the local
 galaxy density = 1 Mpc$^{-2}$.  The solid
 lines represent galaxies with local galaxy density$>1$. The dashed lines represent
 galaxies with local galaxy density$\leq 1$.
 Early-types
 are in the upper left panel. Intermediate-types, early discs and late discs are
 in the upper right, lower left and lower right panel, respectively. 
}
\end{figure}

 To investigate the assertion that the
 morphology--cluster-centric-radius relation is more fundamental than
 the morphology--density relation (Whitmore et al. 1993), we divide the
 morphology-density relation at $R_{vir}=1$, and plot it in Figure
 \ref{fig:md_rad}. The upper left, upper right, lower left and lower
 right panels represent early-type, intermediate-type, early disc and
 late disc galaxies, respectively. In each panel, the solid lines
 use galaxies with $R_{vir}<1$ and the dashed lines use  galaxies with $R_{vir}\geq1$.
 In the figure, there is no significant difference in the fractions of
 early-type, early disc and late disc galaxies between inside and
 outside of $R_{vir}=1$. However, even at low galaxy
 density ($<1$Mpc$^{-2}$) a huge excess
 of intermediate-type galaxies can be seen in the fractions of
 intermediate-type galaxies with $R_{vir}<1$. The existence of the large
 fraction of intermediate-type galaxies at low galaxy density regions
 inside of $R_{vir}=1$ suggest that intermediate-type galaxies can be
 created even at low density regions inside a cluster, and that the cluster
 environment is more fundamental than local galaxy density in creating
 intermediate-type galaxies. 
  It should be noted that Dominguez et
 al. (2001) performed a similar analysis in their Figures 7 and
 8. However, they did not find the difference in elliptical+S0 fractions
 between small cluster-centric radius and large cluster-centric radius
 with low local galaxy density as we found in Figure
 \ref{fig:md_rad}. One possible explanation to this apparent
 discrepancy is that Dominguez et al. (2001) did not separate
 elliptical and S0 galaxies morphologically. It would be interesting to
 investigate whether their sample shows similar results with ours when
 elliptical and S0 galaxies are separated. 
 In Figure \ref{fig:mr_5th}, we plot the
 morphology--cluster-centric-radius relation divided by the local galaxy
 density at 1 Mpc$^{-2}$. The panels and symbols are same as in Figure \ref{fig:md_rad}.
 Since the local galaxy density and the cluster-centric-radius
 correlates themselves, it is difficult to see whether one is more
 fundamental than the other in this figure. However, in the upper right
 panel,  intermediate-type fractions increase toward small
 cluster-centric-radius in both high and low local galaxy density
 regions, supporting our finding in Figure  \ref{fig:md_rad}.

 %

\subsection{Physical Sizes of Galaxies}

\begin{figure}
\includegraphics[scale=0.2]{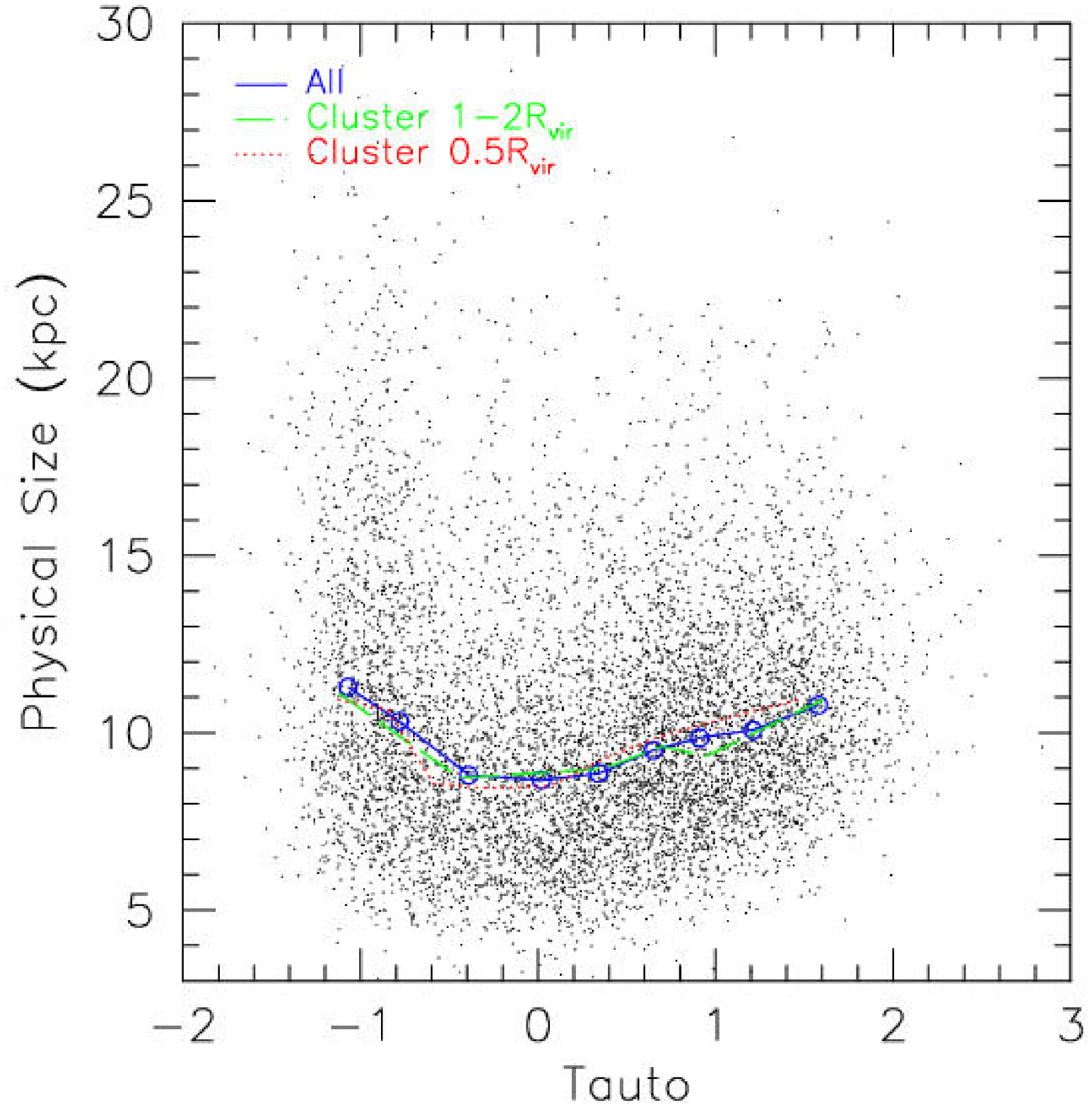}
\includegraphics[scale=0.2]{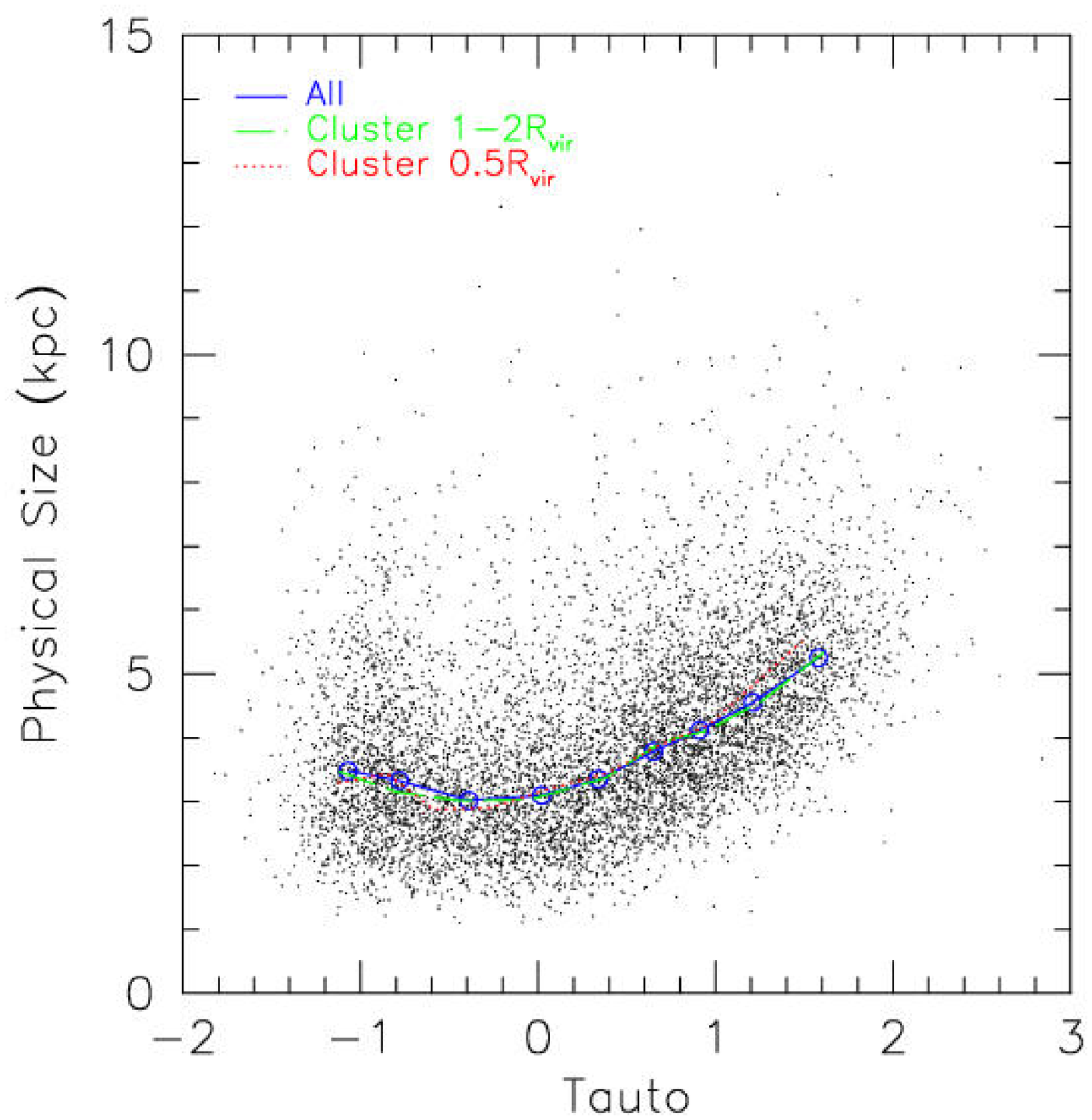}
\caption{
 Physical sizes of all galaxies in the volume limited sample are plotted
 against $Tauto$ with the small dots. Petrosian 90\% and 50\%
 flux radius in $r$ band is used to calculate physical sizes of
 galaxies in the left and right panel, respectively. The solid, dashed
 and dotted  lines
 show medians of all  galaxies, galaxies with 1$<R_{vir}<$2 and galaxies
 with $R_{vir}<$0.5 in the volume limited sample, respectively. In both
 panels, the medians turn over around  
 $Tauto\sim$0, corresponding to S0 population. 
}\label{fig:size} 
\end{figure}

 It is important to understand relative galaxy sizes when we
 discuss transformation of galaxies.
 In the left panel of figure \ref{fig:size}, we plot physical galaxy sizes
 calculated using Petrosian 90\% flux radius in $r$ band, against
 $Tauto$ for all galaxies in the volume limited sample in the small
 dots. The solid, dashed 
 and dotted lines
 show medians of all galaxies, galaxies with 1$<R_{vir}<$2 and galaxies
 with $R_{vir}<$0.5 in the volume limited sample, respectively. All the
 three lines show a consistent trend, suggesting that our $Tauto$
 parameter is not affected by the specific environment of galaxies.
 Above $Tauto$=0,  galaxy sizes decrease with decreasing 
 $Tauto$. However, below $Tauto$=0, galaxy sizes increase with decreasing
 $Tauto$. The same trend can be found in the right panel of  figure
 \ref{fig:size} where we used  Petrosian 50\% flux radius to perform the
 same analysis. 
 We discuss the result in conjunction with
 Figs. \ref{fig:md_ann_ytype} and \ref{fig:mr} in Section \ref{discussion}.

\subsection{Comparison with the MORPHS Data}
\label{Oct 30 18:41:06 2002}

{\scriptsize
\begin{table*} 
\begin{center} 
\caption{\centerline {\sc \label{tab:morphs}
 the MORPHS cluster sample}} 
\vspace{0.1cm}
\begin{tabular}{lccrccccc} 
\hline\hline
\noalign{\smallskip}
{Name} & {R.A.} & {Dec.}  & {$z$} & filter  & {L$_X$ $_{0.3-3.5 keV}(10^{43} h^{-2}$ ergs s$^{-1}${\scriptsize
 )}} & $\sigma$ (km s$^{-1}$)\hfil  \cr
\noalign{\smallskip}
\hline
\noalign{\smallskip}
A370\#2 & 02~40~01.1 & $-$01~36~45 &  0.37 & F814W &  2.73&1350 [34]  \cr
Cl1447+23 & 14~49~28.2 & $+$26~07~57 &  0.37 & F702W &  ...\hfil & ... \cr
Cl0024+16 & 00~26~35.6 & $+$17~09~43 &  0.39 & F814W & 0.55 & 1339 [33]  \cr
Cl0939+47 & 09~43~02.6 & $+$46~58~57 &  0.41 & F814W &  1.05 & 1081 [31]  \cr
Cl0939+47\#2 & 09~43~02.5 & $+$46~56~07 &  0.41 & F814W & 1.05 & 1081[31]  \cr
Cl0303+17 & 03~06~15.9 & $+$17~19~17 & 0.42 & F702W & 1.05 & 1079 [21]  \cr
3C295 & 14~11~19.5 & $+$52~12~21 &  0.46 & F702W & 3.20 & 1670 [21]  \cr
Cl0412$-$65 & 04~12~51.7 & $-$65~50~17 &  0.51 & F814W & 0.08 & ...\hfil  \cr 
Cl1601+42 & 16~03~10.6 & $+$42~45~35 &  0.54 & F702W & 0.35 & 1166 [27]  \cr
Cl0016+16 & 00~18~33.6 & $+$16~25~46 &  0.55 & F814W & 5.88 & 1703 [30]  \cr
Cl0054$-$27 & 00~56~54.6 & $-$27~40~31 & 0.56 & F814W &  0.25 & ...\hfil \cr
\noalign{\hrule}
\end{tabular}
\end{center} 
\end{table*}
}

\begin{figure}
\includegraphics[scale=0.4]{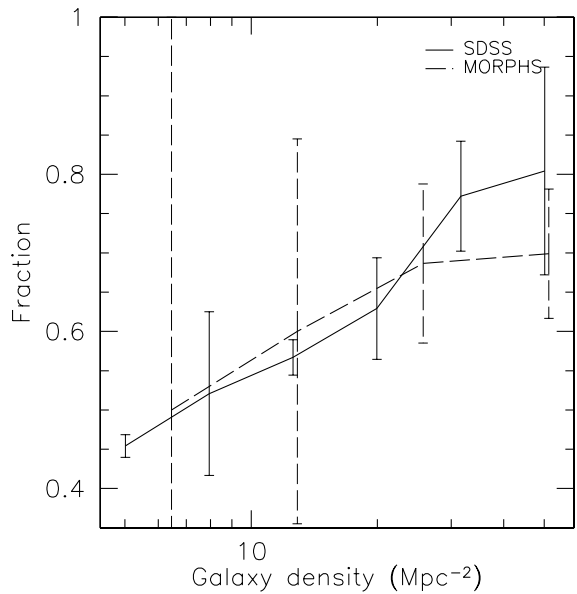}
\caption{
\label{fig:md_morphs} 
 Comparison of the morphology-density relations of the SDSS (low redshift)
 and the MORPHS (high redshift). Fraction of early-type galaxies are
 plotted against local galaxy density within 250 kpc (Note that the local
 galaxy density here is measured in a different way from that in previous sections). The MORPHS data are plotted in the solid line, and the SDSS data are plotted in the dashed line.
}
\end{figure}

 In this section, we compare the morphology-density relation of the SDSS
 data ($z\sim$0.05) with that of the MORPHS data ($z\sim$0.5).
 The MORPHS data are used in Dressler et al. (1997) to study
 the morphology-density relation in high redshift clusters,
 and publicly available in Smail et al. (1997). The data consist of 10 rich
 galaxy clusters at a redshift range $z=$0.37-0.55 as summarized in
 table \ref{tab:morphs}. The sharp imaging
 ability of the Hubble Space Telescope made it possible to measure galaxy
 morphology at this further away in the universe. We use concentration
 parameter given in Smail et al. (1997) as an automated morphology of
 the sample. It is more interesting to apply the $Tauto$ parameter to
 the MORPHS data. However, any morphological
 parameter needs careful calibration between high and low redshift
 imaging data where a pixel size and a psf size compared with a galaxy size
 are very different. We found such calibration is easier and more
 trustable for a  concentration parameter as is used below.
 As for the galaxy density, we count a number of galaxies
 brighter than $Mr^*=-$19.0 within 250 kpc and subtracted average galaxy
 number count of the area (Glazebrook et al. 1995).  Magnitude (either
 F702W or F814W) are
 $k$-corrected and transformed to the SDSS $r$ band using the relation
 given in Fukugita et al. (1995). 
 To make as fair comparison as
 possible, we re-measured the SDSS morphology-density relation using
 as similar criteria as possible. We re-measure galaxy density by
 counting galaxies within
 250 kpc and $\pm$1000 km s$^{-1}$ and brighter than $Mr^*=-$19.0 in the SDSS
 data (0.01$<z<$0.054). The number of 
 galaxies is divided by the size of the area (250$^2$ $\pi$ kpc$^2$ if it does
 not go outside of the boundary. If it does, the area is corrected
 accordingly.) 
 We also match the criteria for both of concentration parameters. The
 concentration parameter of the SDSS is measured as a ratio of
 Petrosian 50\% light radius to 90\% light radius. The concentration
 parameter of the  
 MORPHS data is measured using the Source Extractor. Furthermore, the
 seeing size compared with typical galaxy size is not exactly the same
 between two samples. Therefore, we have to calibrate these two
 concentration parameters. 
 Fortunately part of the SDSS galaxies are morphologically classified by
 eye in Dressler (1980). Since the MORPHS data are eye-classified by the same
 authors (Dressler et al. 1997), we regard these two eye-classified
 morphology essentially the same and use them to calibrate two
 concentration parameters. If we use the SDSS concentration criteria, $Cin<$0.4,
 It leaves 76\% of eye-classified elliptical galaxies (24\% contamination). By
 adjusting concentration parameter for the MORPHS data to this value, we
 found that the MORPHS concentration parameter of 0.45 leaves 75\% of
 eye-classified elliptical galaxies. We regard these essentially the same
 criteria and adopt 0.45 as a criteria for the MORPHS concentration
 parameter which corresponds to that of the SDSS($Cin$=0.4). In figure
 \ref{fig:md_morphs}, we plot fractions of early-type galaxies against
 galaxy density for both the SDSS and MORPHS data in the solid and
 dashed lines, 
 respectively. Quite interestingly, two morphology-density relations lie
 on top of each other. Since the MORPHS data only exist for cluster
 region, we are not able to probe into as low density regions as in
 figure \ref{fig:md_ann_ytype}. However, it shows that the
 morphology-density relation was already established at $z\sim$0.5 .
 There is a sign of slight excess of early-type galaxies in the SDSS
 data in two  highest density bins.

\section{Discussion}\label{discussion}

\subsection{Early-type Fractions}
 In previous section, we have presented fractions of early-type galaxies
 in several different ways. Since our morphological classification,
 density measurement are different from most of previous work, it is
 important to know how these early-type fractions differ due to the
 choice of relevant parameters. We use the result of Whitmore et
 al. (1993) as a benchmark for our study since they applied various
 systematic corrections carefully 
 and the results are relatively widely used. Their fraction ratio of
 galaxies is elliptical:S0:spiral=18\%:23\%:59\%. 
   In figure \ref{fig:ann_cin}, we have
 55\% of early-type galaxies in the least dense bin and 85\% of them in the
 densest bin.    In figure \ref{fig:mr_cin}, our early-type fractions vary from 60\% to
 90\%. Between these two figures, our values agree each other within the
 errors, suggesting
 our values are internally consistent. However, our early-type fractions
 are slightly lower than the sum of ellipticals and S0s (41\%) in
 Whitmore et al. (1993). As noted in section \ref{analysis}, this comes
 from our choice of $Cin$=0.4 
 criteria slightly leaned toward spiral galaxies. Figure
 \ref{fig:ann_cin} showed that slight change in $Cin$ criteria can
 change absolute amount of early-type galaxies dramatically, and thus,
 careful attention is needed when comparing our work with others. In our
 case, $Cin$=0.37 criteria shown in dotted line in figure
 \ref{fig:ann_cin} is closer to the classification of Whitmore et al. (1993).

   In figure \ref{fig:md_ann_ytype}, we have 10\% of
 early-type in the least dense bin and 25\% in the densest
 bin. Intermediate-type galaxies are
 25\% in the least dense bin.
   In figure \ref{fig:mr}, early-type fractions  vary from 15\% to
 30\% and intermediate-type fractions vary from 25\% to 40\%. 
 Again our values
 are consistent within the errors internally. In addition, in these cases,
 values are only two sigma away from Whitmore et al. (1993). If we add
 early-type fractions and intermediate-type fractions, our values are 30\%
 and 35\%, whereas Whitmore 
 reports 41\%. Therefore, in case of $Tauto$ parameter, our choice of
 criteria is similar to that of Whitmore et al. (1993).

\subsection{The Morphology-Density Relation with $Cin$}
 
 In section \ref{morphology_density}, two interesting results are
 found in the SDSS data using $Cin$=0.4 as a classification criterion;
   (i) the morphology-density relation exists in the SDSS data,
 however it flatters out at low density. (ii) the characteristic galaxy
 density is at  around 3 Mpc$^{-2}$. In section \ref{Oct 30 18:22:01
 2002}, we analyzed the
 morphology-density relation in the view point of the morphology-radius
 relation. The flattering at low density is seen as well with its turning
 point at around 1 $R_{vir}$. 
  Since computer simulations sometimes use different magnitude range,
 different density measurement and different morphological
 classification (often bulge-to-disc ratio), it is difficult to do
 accurate direct comparison.
 However, both of morphology-density relation and morphology-radius relation are
 qualitatively in agreement with computer simulations such as Okamoto et
 al. (2001), Diaferio et al. (2001), Springel et al. (2001) and Benson
 et al. (2001).
  The flattering of morphology-density relation we saw in both figures
 \ref{fig:ann_cin} and \ref{fig:mr_cin} is
 interesting since it suggests that whatever physical mechanism is
 responsible for morphological change of late-disc galaxies in going toward
 dense regions, the mechanism starts working at galaxy density  $\sim$3
 Mpc$^{-2}$ or higher.

 Various mechanism can be responsible for the
 morphological change. 
  These include ram-pressure stripping of gas (Spitzer \& Baade 1951;
 Gunn \& Gott 1972; Farouki \& Shapiro 
 1980; Kent 1981; Abadi, Moore \& Bower 1999; Fujita \& Nagashima 1999;
 Quilis, Moore \& Bower 2000), 
 galaxy harassment (Moore et al. 1996, 1999;  Fujita 1998), cluster tidal forces (Byrd
 \& Valtonen 1990; Valluri 1993;  Fujita 1998),  enhanced star formation (Dressler \&
 Gunn 1992), and   removal \& consumption of the gas (Larson,
 Tinsley \& Caldwell 1980; Bekki et al. 2002).
  It is yet unknown exactly what processes play major roles in creating
 morphology-density relation. However, the mechanism must be the one that
 works at galaxy density 3 Mpc$^{-2}$ or higher. It is also interesting
 to note that this characteristic density coincides with the density
 where galaxy SFR abruptly drops (Lewis et al. 2002;
 Gomez et al. 2003). The coincidence suggests that the same mechanism
 might be
 responsible for both morphology-density relation and the truncation of star
 formation.

%

\subsection{The Morphology-Density Relation with $Tauto$}
 In figures \ref{fig:md_ann_ytype} and \ref{fig:mr}, we further studied
 the morphology-density and the morphology-radius relation using $Tauto$
 parameter, which allows us to divide galaxies into four categories
 (early, intermediate, early-disc and late-disc). 
  In addition to the general trend found in the previous
 section, we found two characteristic changes in the relation at around
 galaxy density 1 and 6 Mpc$^{-2}$ or in terms of radius, 0.3 and 1
 virial radii. In the sparsest regions (below 1 Mpc$^{-2}$ or outside of
 1 virial radius), both 
 relations becomes almost flat, suggesting the responsible physical
 mechanisms do not  work very well in these regions. In the intermediate
 regions (density between 1 and 6 Mpc$^{-2}$ or virial radius between
 0.3 and 1), intermediate-type fractions dramatically increase toward denser or smaller
 radius regions, whereas  fractions of late-disc galaxies
 decrease. In the densest regions (above 6 Mpc$^{-2}$ or inside of
 0.3 virial radius), interestingly intermediate-type fractions decrease, and in turn,
 early-type fractions radically increase suddenly. The change in the
 densest region are further confirmed in figures \ref{fig:md_es0} and
 \ref{fig:mr_es0}, where we plotted intermediate to early-type number ratio as a
 function of density or cluster-centric-radius. In both figures,
 intermediate to early-type ratio declines suddenly at the densest region. 

  The existence of two characteristic change in both the morphology-density
 and the morphology-radius relation suggests the existence of two
 different physical mechanisms responsible for each morphological fraction
 change. 
   In the intermediate region (density between 1 and 6 Mpc$^{-2}$
 or virial radius between 0.3 and 1), the mechanism creates intermediate-type galaxies
 mostly, by reducing fractions of late-disc galaxies. Although there is not much
 change in early-disc fractions, perhaps it is natural to imagine that the
 mechanism turns late-disc galaxies into early-disc galaxies, and then
 early-disc galaxies
 into intermediate-type galaxies. As figure \ref{fig:size} shows, median sizes of galaxies
 gradually declines from late-discs to intermediate-type galaxies, suggesting calm, gradual
 transformation of galaxies, maybe due to the truncation of star
 formation as observed at the same environment by Gomez et al. (2003)
 and Lewis et al. (2002). After the truncation of star formation, outer
 part of a galaxy disc fades 
 away as massive stars die. In fact, such a galaxy in transition is
 found by Goto et al. (2003a), who showed that spiral galaxies with no
 emission lines (passive spirals) preferentially live in cluster
 infalling regions.
 The calm, quiescent truncation of the star formation is also supported
 by our findings in Section \ref{density_radius}, where we found that
 intermediate-type galaxies can be created inside of $R_{vir}=1$
 regardless of the local galaxy density. The results imply that cluster
 environment such as interaction with hot, plasma gas associated with cluster potential
 is more important to create intermediate-type galaxies  than the process
 induced by the enhancement of local 
 galaxy density (e.g., galaxy-galaxy merger/interaction).
 The plausible
 candidates of the responsible mechanism includes ram-pressure stripping (Gunn \& Gott 1972; Farouki
 \& Shapiro 1980; Kent 1981; Fujita \& Nagashima 1999;   Abadi, Moore \&
 Bower 1999; Quilis, Moore \& Bower 
 2000; Fujita 2001, 2003), unequal mass galaxy mergers (Bekki et
 al. 1998), galaxy
 harassment (Moore et al. 1999)   and  truncation of
 star formation due to the cluster environment (strangulation; Larson, Tinsley \&
 Caldwell 1980; Bekki et al. 2002; Mo \& Mao 2002; Oh \&
	Benson 2001), evaporation of the cold gas in disc galaxies via
	heat conduction from the surrounding hot ICM (Cowie \& Songaila
	1977; Fujita 2003).
       Although several authors indicated that these regions are too low
 gas density for stripping to happen (e.g., Lewis et al. 2002), 
	Fujita (2003) pointed out that ram-pressure stripping
 may be effective in cluster sub-clump regions. In fact, Kodama et
 al. (2001) found galaxy colours suddenly change in such sub-clump
 regions.
 Perhaps, it is also worth noting that E+A (K+A or post-starburst)
 galaxies often thought to be cluster-related
 are found to have their origin in merger/interaction with accompanying
 galaxies (Goto et al. 2003b,c), and thus E+A galaxies are not likely to be
 a product of the morphological transition in these cluster regions.

  Very different consequences are found in the densest region
 (above 6 Mpc$^{-2}$ or inside of 
 0.3 virial radius), where the mechanism decrease intermediate-type
 fractions and increase 
 early-type fractions. In figure \ref{fig:size}, there is a significant
 increase in median galaxy sizes from intermediate-type to early-type
 galaxies. Both of these 
 observational results suggest a very different mechanism from
 intermediate region is working in the densest region. Since galaxy size
 becomes larger from intermediate-type to early-type galaxies (Figure \ref{fig:size}),
 merging scenario is one of the 
 candidate mechanisms. 
   Computer simulations based on the galaxy merging
 scenario reported the deficit of intermediate-type galaxies (Okamoto et al. 2001;
 Diaferio et al. 2001; Springel et al. 2001; Benson 
 et al. 2001), which we might have seen
 observationally in the densest region of our data. 
   However, theoretical work in the literature suggests that
 merging/interaction is difficult to 
  happen in cluster core regions since relative velocity of galaxies are too
  high in such regions (Ostriker 1980; Binney  \& Tremaine 1987; Mamon
 1992; Makino \& Hut 1997; but see also Mihos 2003). If this is the case, 
 dominance of old early-type galaxies in cluster cores might be so extreme
 that early-type fractions overwhelm the increase of intermediate-type fractions in
 cluster cores.
 In previous work, Postman \& Geller (1984) pointed out that their
 morphology-density relation has two breaks.
 Dominguez et
 al. (2002) suggested that there are two mechanisms in the
 morphology-density relation; one with global nature and the other with
 local effects. 
  Our findings of two characteristic changes in the
 morphology-density relation is perhaps an observational result of the same
 physical phenomena as these from a different point of view.

\subsection{Comparison with the MORPHS Data}  
  In section \ref{Oct 30 18:41:06 2002}, we compared the
  local morphology-density relation (SDSS; $z\sim$0.05) with that of higher
  redshift (MORPHS; 0.37$<z<$0.5). Interestingly, two
  morphology-density relations agreed each other. The agreement suggests
  that morphology-density relation was already established at $z\sim$0.5
  as it is in the present universe, i.e., the origin of
  morphology-density relation stays much higher redshift
  universe.  In the densest environments, there might be a sign of
  excess early-type fractions in the SDSS than in the MORPHS. Although
  two data points agrees within the error, such an excess of early-type
  galaxies might suggest additional formation of elliptical/S0 galaxies
  between $z=$0.5 and $z=$0.05. If it is the case, the result may be consistent
  with Dressler et al. (1997) and Fasano et al. (2000), where they
  observed the increase of S0 galaxies toward lower redshift and proposed
  spiral to S0 transformation.
  Little evolution of morphology-density relation is also
  interesting in terms of comparison with computer simulations. 
  Benson et al. (2001) predicted that the evolution of
  morphology-density relation will be seen as a shift in the decrease
  of early-type fractions without significant change in slope. 

  However, a
  caveat is that absolute value in Figure \ref{fig:md_morphs}
  depends solely on the calibration of concentration parameters of the
  both data, which by nature is difficult to calibrate accurately due to
  the large scatter in both concentration parameters. Therefore, the
  results on evolution should not be over-interpreted. In addition, the
  richness of the MORPHS clusters and the SDSS clusters are very different.
  The MORPHS sample consists of rich clusters with their velocity
  dispersion $>1000$ km s$^{-1}$ (Table \ref{tab:morphs}), whereas the
   median velocity  dispersion of the SDSS clusters is $\sim$700  km
  s$^{-1}$ (Table 1 of Gomez et al. 2003). Since it is often reported that
  rich clusters might have fewer amount of blue/late-type galaxies than
  poor clusters (Margoniner et al. 2001; Goto et al. 2003d), it is more
  ideal to use clusters with similar richness for both high and low
  redshift samples. Such a study will become possible in the near future
  as the SDSS observes larger volume to find many rich clusters.


%
%
%
%
%

\section{Conclusions}\label{conclusion}

 We have studied the morphology-density relation and the
 morphology--cluster-centric-radius relation using a volume limited
 sample of the SDSS
 data (0.05$<z<$0.1, $Mr^*<-$20.5). Compared with previous studies,
 major improvements in this work are;
 (i) automated  galaxy morphology, (ii) three dimensional local galaxy density
 estimation, (iii) the extension of the morphology-density relation into
 the field region. 
 Our findings can be summarized as follows.
   
 Both the morphology-density relation and the
 morphology--cluster-centric-radius relation are seen in the SDSS data for both of our
 automated morphological classifiers, $Cin$ and
 $Tauto$. 

  We found there are two characteristic changes in both the
 morphology-density and the morphology-radius relations, suggesting two
 different mechanisms are responsible for the relations.
  In the sparsest regions (below 1 Mpc$^{-2}$ or outside of 1 virial
 radius), both relations  become less effective, suggesting the
 responsible physical mechanisms require denser environment. The
 characteristic density or radius 
 coincides with the sharp turn in SFR-density relation
 (Lewis et al. 2002; Gomez et al. 2003), suggesting the same mechanism
 might be responsible for both the morphology-density relation and SFR-density relation. 
   In the intermediate density regions, (density between 1 and 6
 Mpc$^{-2}$ or virial radius between 0.3 and 1), intermediate-type fractions increase
 toward denser regions, whereas late-disc fractions
 decrease. Considering the median size of intermediate-type galaxies are smaller than
 that of late-disc galaxies
 (figure \ref{fig:size}) and SFR radically declines in
 these regions, the mechanism
 that gradually reduces star formation might be responsible for
 morphological changes in these intermediate density regions
 (e.g., ram-pressure stripping). Quiescent truncation of star formation
 is also supported by our findings in Section \ref{density_radius},
 where we found that  intermediate-type galaxies can be created inside of $R_{vir}=1$
 regardless of the local galaxy density. The results imply that cluster
 environment such as interaction with plasma gas associated with cluster potential
 is more important to create intermediate-type galaxies  than the process
 induced by the enhancement of local 
 galaxy density (e.g., galaxy-galaxy merger/interaction).
 The mechanism is likely to stop star formation in 
 late-disc galaxies, then late-disc galaxies becomes early-disc galaxies
 and eventually turns into smaller intermediate-type galaxies after their outer discs and
 spiral arms become  invisible as stars die. 
   In the densest regions (above 6 Mpc$^{-2}$ or inside of
 0.3 virial radius), intermediate-type fractions decreases radically and early-type
 fractions increase. This is a contrasting results to that in
 intermediate regions and it suggests that yet another mechanism is
 responsible for morphological change in these regions. Considering that
 elliptical galaxies can be observed in high redshift universe (e.g., van
 Dokkum et al. 2000), dominance of elliptical galaxies might be so
 extreme in cluster cores
 that early-type fractions overwhelm the increase of intermediate-type fractions.


%
%
%

 We also compared our morphology-density relation from the SDSS
 ($z\sim$0.05) with that of the MORPHS data ($z\sim$0.5). Two
 relations lie on top of each other, suggesting that the
 morphology-density relation was already established at $z\sim$0.5 as is
 in the present universe. In the densest bin, a slight sign of excess
 elliptical/S0 fraction was seen in the SDSS data, which might be
 indicating the formation of additional elliptical/S0 galaxies between $z=$0.5 and
 $z=$0.05.

\section*{Acknowledgments}

 We are grateful to Michael L. Balogh, Robert C. Nichol, Masami Ouchi,
 Masayuki Tanaka, Alex Finoguenov and Ann Zabludoff for useful discussions.
 We wish to express our gratitude to Takashi Okamoto, Naoyuki Tamura,
 Masahiro Nagashima for useful conversation.  We thank the anonymous
 referee for many insightful comments. 
 T. G. acknowledges financial support from the Japan Society for the
 Promotion of Science (JSPS) through JSPS Research Fellowships for Young Scientists.

 The Sloan Digital Sky Survey (SDSS) is a joint project of The
University of Chicago, Fermilab, the Institute for Advanced Study, the
Japan Participation Group, the Johns Hopkins University, the
Max-Planck-Institute for Astronomy, New Mexico State University,
Princeton University, the United States Naval Observatory, and the
University of Washington. Apache Point Observatory, site of the SDSS
telescopes, is operated by the Astrophysical Research Consortium
(ARC).  Funding for the project has been provided by the Alfred
aP.~Sloan Foundation, the SDSS member institutions, the National
Aeronautics and Space Administration, the National Science Foundation,
the U.~S.~Department of Energy, Monbusho, and the Max Planck
Society. The SDSS Web site is http://www.sdss.org/.


\label{lastpage}

\end{document}